\documentclass[%
 reprint,
 amsmath,amssymb,
 aps,
]{revtex4-2}
\usepackage{mathrsfs}
\usepackage{color}
\usepackage{ulem}
\usepackage{graphicx}
\usepackage{dcolumn}
\usepackage{bm}

\newcommand{\tb}[1]{\textcolor{black}{#1}}

\begin{document}

\preprint{APS/123-QED}

\title{Interaction-induced quantum spin Hall \tb{insulator} in the organic Dirac electron system $\alpha$-(BEDT-TSeF)$_2$I$_3$}

\author{Daigo Ohki$^1$}
\email{dohki@s.phys.nagoya-u.ac.jp}
\author{Kazuyoshi Yoshimi$^2$}%
\author{Akito Kobayashi$^1$}
 \affiliation{$^1$Department of Physics, Nagoya University, Furo-cho, Chikusa-ku, Nagoya, 464-8602 Japan \\
  $^2$Institute for Solid State Physics, University of Tokyo, Chiba 277-8581, Japan \\
}%




\date{\today}

\newcommand{\ohki}[1]{\textcolor{black}{#1}}

\begin{abstract}
Focusing on the recently-discovered candidate topological insulator $\alpha$-(BEDT-TSeF)$_2$I$_3$ -- having two-dimensional charge-neutral Dirac cones in a low symmetry lattice -- we combine ab-initio and extended-Hubbard model calculations to deal with spin-orbit and non-local repulsive interactions, and find a realization of an interaction-induced quantum spin Hall (QSH) insulator, similar to the one proposed in the honeycomb lattice under next-nearest neighbor repulsions.
In the absence of repulsive interactions, a topological insulator appears by the spin-orbit coupling and is characterized by a nonzero spin Chern number.
By considering up to next-nearest neighbor repulsions at Hartree-Fock level, the intrinsic spin-orbit gap is found to grow by orders of magnitude and a QSH \tb{insulating phase} appears that has both a finite spin Chern number and order parameter.
Transport coefficients and spin susceptibility are calculated and found to consistently account for most of the experimental findings, including the metal-to-insulator crossover occurring at $\sim50$ K as well as the Berry phase change from 0 to $\pi$ under hydrostatic pressure.
We argue that such a QSH \tb{insulating phase} does not necessitate a sizeable spin-orbit interaction to generate a large insulating gap, which is highly advantageous for the search of novel topological phases in generic materials having low symmetry lattice and/or small spin-orbit coupling.
\end{abstract}

\maketitle


\section{Introduction}

In condensed-matter physics, the studies of pseudo-relativistic Dirac electrons in solids -- such as massless Dirac fermions in graphene \cite{Wallace, Novoselov} and organic conductors \cite{Kajita1992, Tajima2000, Kobayashi2004, Katayama2006, Kobayashi2007, Goerbig2008, Kajita2014, Tajima2006}, and massive Dirac fermions in bismuth \cite{Wolff, Fukuyama1970} -- have attracted great attention because of the anomalous electronic excitation and correlation properties, including quantum conduction linked to a universal conductivity \cite{Shon}, gigantic diamagnetism \cite{Fukuyama1970}, or logarithmic self-energy corrections due to unscreened Coulomb interaction \cite{Kotov, HirataScience}.
Nontrivial features of the electron wavefunctions provide an important addition to such studies, leading to a rich variety of new states of matter with intriguing momentum-space topology that is robust against local perturbations.
For example, when time-reversal symmetry is broken without external magnetic field, a nontrivial state with the quantum anomalous Hall (QAH) effect \cite{Haldane} is to be realized; This state has an insulting bulk gap and chiral edge states while the translational symmetry is preserved, as has been shown by Haldane for a tight-binding model defined on a honeycomb lattice with next-nearest-neighbor hoppings and staggered flux.
In the presence of time-reversal symmetry and spin-orbit coupling (SOC), Kane and Mele further proposed the quantum spin Hall (QSH) insulator \cite{KaneMele, HasanKane, Ando} with broken $SU(2)$ spin symmetry by generalizing Haldane's model.
Just like in the case of QAH effect, the QSH insulator has a bulk gap while it shows counter-propagating helical edge states, which is more recently known as a prototype of time-reversal symmetry protected topological insulator (TI).

Notably, even in the absence of SOC, it is known that both QAH and QSH \tb{insulating phases} with broken time-reversal symmetry and $SU(2)$ spin symmetry can be generated by repulsive interactions.
These states were first proposed by Raghu et al. \cite{Raghu} by using an extended Hubbard model for a honeycomb lattice and considering up to the next-nearest-neighbor interactions that are frustrated within the same bipartite sublattice.
However, more recent density matrix renormalization-group studies failed to uncover such interaction-induced phases for realistic parameters \cite{Rachel,Wen2010,Weeks,Dauphin,Noel2013,Grushin2013,Daghofer2014,Duric,Capponi,Motruk,Tianhan,Venderbos}, proving their realization rather elusive in honeycomb lattice.
Nonetheless, the idea of QSH phase has been shown to be applicable to wider range of physical systems than originally expected; It leads to the paradigm-shift concept of symmetry-protected topological phases \cite{Rachel}, which is theoretically applicable to wider classes of materials from heterostructures of transition-metal oxides \cite{Xiao2011, Ruegg2011, Ruegg2012} and dichalcogenides \cite{Qian} to element pnictogens \cite{Fukui} to silicene \cite{Ezawa}, and experimentally confirmed in HgTe/CdTe quantum wells \cite{Bernevig}.
Given such widespread interest in searching novel topological phases, Raghu's original construct of interaction-induced phases -- although not feasible in graphene -- would still have a unique significance and versatility.
This is especially because interactions may not necessarily require strong SOC in stabilizing such phases \cite{Rachel} and would, therefore, help the phases to be stabilized in more generic conditions, including systems having low-lattice symmetry and/or weak SOC.
However, investigation and elucidation of interaction-induced topological phases is greatly limited to honeycomb lattice, and their relation to repulsive and spin-orbit interactions in more generic lattices remains highly unclear.

In this paper, to shed new light on the physics of interacting topological phases in general low-symmetry lattices, we combine an extended-Hubbard model with density functional calculations to simultaneously consider non-local repulsive and intrinsic spin-orbit interactions. We apply this scheme to the newly-discovered, TI candidate organic conductor $\alpha$-(BEDT-TSeF)$_2$I$_3$ ($\alpha$-(BETS)$_2$I$_3$) [where BEDTTSeF (BETS) is bis(ethylenedithio)tetraselenafulvalene], which in the absence of SOC has tilted massless Dirac cones that are protected by inversion symmetry.
By constructing and solving effective mean-field theory, we propose a realization of an interaction-induced QSH insulator that is characterized not only by a topological invariant but also with a distinctive order parameter.
In the absence of repulsive interactions, the SOC opens a finite gap at the band-crossing point that is located at the Fermi energy $E_F$, making the system to be a TI with helical edge states.
In the presence of nearest and next-nearest neighbor repulsions while in the absence of SOC, the system becomes either interaction-induced QAH or QSH insulators (degenerate at Hartree-Fock level) that possess a finite Chern number together with an order parameter
[Here, the term "TI" will be exclusively used to describe a time-reversal protected insulator with helical states that is solely induced by SOC, following the original sense of Kane and Mele \cite{KaneMele}.
The "QSH" state will be distinguished from TI in such a way that it describes an insulating phase with helical states that is predominantly generated by repulsive interactions regardless of the size of SOC, following the spirit of Raghu et al. \cite{Raghu}.].
By employing the values of spin-orbit and repulsive interactions estimated from the first-principles studies, we find the interactions to stabilize a QSH insulator with a nonzero spin Chern number, in which the repulsions greatly enhance the intrinsic spin-orbit gap and generate a finite order parameter.
Conductivity within Kubo formula and dynamic spin susceptibility linked to the nuclear spin-lattice relaxation rate $1/T_1$ were calculated and found to reasonably account for most of the experimental findings reported by transport \cite{Inokuchi, Kawasugi, TajimaPriv} and nuclear magnetic resonance (NMR) \cite{Hiraki, Fujiyama} measurements.
Further, combined with temperature ($T$) dependence analysis of mean-field theory, we find that the QSH \tb{insulating phase} continuously enhances the underlying spin-orbit gap toward lower $T$, which seemingly plays an essential role in explaining the observed sharp increase of resistivity developing below 50 K \cite{Inokuchi, Kawasugi}.

The rest of the paper is organized as follows.
In Section II, we provide the background of the organic-conductor family that we focus here and introduce the model descriptions for such systems that host tilted Dirac cones defined in a quasi-two-dimensional (quasi-2D) lattice.
It also offers the mathematical frameworks employed in our density functional calculations and mean-field theory that are constructed on an extended Hubbard model, along with those used in the calculations of conductivity and spin susceptibility.
In Section III, we show how the nearest and next-nearest neighbor interactions induce and stabilize a QSH insulator for realistic values of hopping integrals and spin-orbit interactions estimated from ab-initio calculations, and see how such QSH state can be relevant to the understanding of transport and NMR observations.
Finally, Sec. IV gives the conclusion and discussions.

\section{Background and Formulation}
%
\subsection{Background of the organic conductor $\alpha$-(BEDT-TSeF)$_2$I$_3$}
To study the impacts of repulsive interactions on stabilizing topological states in low symmetry lattices, we focus on the organic conductor $\alpha$-(BETS)$_2$I$_3$ and employ it as a model system, where in the absence of spin-orbit interaction, a pair of band-crossing points with a tilted Dirac cone dispersion appear and are protected by inversion symmetry.
This material has a layered structure similar to that in the more-studied relative $\alpha$-(BEDT-TTF)$_2$I$_3$ ($\alpha$-(ET)$_2$I$_3$), where BEDT-TTF (ET) is bis(ethylenedithio)-
tetrathiafulvalene \cite{Kajita1992, Tajima2000, Kobayashi2004, Katayama2006, Kobayashi2007, Goerbig2008, Kajita2014, Tajima2006, KitouSawaTsumuraya, TsumurayaSuzumura, SuzumuraTsumuraya}.
(Here, BETS molecules are derived from ET molecules by replacing the underlying sulfur atoms with selenium atoms.)
In both systems because the inter-layer hopping integrals are significantly smaller than the in-plane ones, the inter-layer coupling can be omitted as a first approximation, and one can reasonably describe the systems with a 2D lattice model \cite{Kino}.
Further, both systems have four molecules in the unit cell where three of them are crystallographically nonequivalent [dubbed A (=A$'$), B, and C in Fig. \ref{Fig:Wannier-Fit}(a)] \cite{KitouSawaTsumuraya}.
By neglecting small inter-layer hopping integrals, the overlaps of wavefunctions between neighboring molecules in each layer generate four 2D energy bands near $E_F$ as in Fig. \ref{Fig:Wannier-Fit}(b), and lead to two Dirac cones locating at general incommensurate wavenumbers in the first Brillouin zone, with their crossing points fixed at $E_F$ because of the $3/4$-filling of the energy band.

In the well-studied material $\alpha$-(ET)$_2$I$_3$, experimental works found a prototypical first-order metal-insulator transition at $T = 135$ K with a stripe charge order and inversion symmetry breaking probably linked to nearest-neighbor repulsive interactions \cite{Seo2000, Takahashi, Kakiuchi} that cause exotic charge and spin responses \cite{TanakaOgata, Ishikawa, Beyer, Liu, Ohki2019}.
Application of hydrostatic pressure ($P$) suppresses the transition and a massless Dirac electron phase appears above $P\sim12$ kbar, in which notable correlation effects of massless electrons have been reported, such as ferrimagnetic spin polarization, velocity renormalization, and
excitonic spin fluctuations \cite{Kobayashi2013, HirataNat2016, Matsuno2017, Matsuno2018, HirataScience, Ohki2020}.
In $\alpha$-(BETS)$_2$I$_3$, experiments found a similar change in the electronic property from metallic to insulating upon cooling, while this change happens more gradually around $T=50$ K via a crossover \cite{Inokuchi, Kawasugi, TajimaPriv}.
Notably, in contrast to $\alpha$-(ET)$_2$I$_3$ X-ray diffraction experiments below this crossover $T$ found no evidence for inversion symmetry breaking and lattice modulation \cite{KitouSawaTsumuraya}.
NMR studies further reported linear and quadratic $T$-dependence across 50 K in the spin susceptibility and the spin-lattice relaxation rate $1/T_1T$, respectively, without any signature of magnetic order \cite{Hiraki,Fujiyama}.
These findings suggest that the crossover is not accompanied by charge, bond, and spin orders.
Upon applying $P$, moreover, the crossover is weakened and vanishes at $P_C\sim5$ kbar \cite{Kawasugi, Kawasugi2}; On the verge of this $P_C$, an interesting remark was recently made by quantum oscillation measurements in hole-doped thin-film samples, which reported an abrupt $P$-induced change in the Onsager phase factor that is translated into a jump of the Berry phase from 0 to $\pi$ at $P_C$ \cite{Kawasugi}.

Theoretically, the difference of the ground states in two systems has been believed to be linked to the different nature and consequence of SOC in these systems because the size of the intrinsic SOC would be quite contrasting between the underlying selenium atoms (in BETS molecules) and sulfur atoms (in ET molecules) \cite{KitouSawaTsumuraya, SuzumuraTsumuraya, Winter}.
Indeed, perturbative calculations found a SOC of the size of 5-10 meV in $\alpha$-(BETS)$_2$I$_3$, whereas it is just 1-2 meV in $\alpha$-(ET)$_2$I$_3$ \cite{Winter}.
First-principles calculations in $\alpha$-(BETS)$_2$I$_3$ using generalized gradient approximation (GGA) \cite{TsumurayaSuzumura} further suggested that a SOC induced gap of an approximate size of 2 meV opened at $E_F$ \cite{Kondo, Alemany, MorinariSuzumura}, making the system to be a TI with helical edge states \cite{TsumurayaSuzumura, SuzumuraTsumuraya, Winter}.
More recently, the influence of spin-orbit interaction was incorporated into tight-binding model by adding correction terms to the transfer integrals, which led to the finding that such SOC gap can partly account for the observed crossover showing a sharp increase of resistivity below 50 K \cite{TsumurayaSuzumura}.
In our previous study, we examined the influence of on-site repulsive interactions by performing calculations based on a 2D Kane-Mele-Hubbard type model \cite{Ohki2020BETS}; There, we found within Hartree approximation that a spin-ordered massive Dirac electron phase was stabilized as a possible ground state, which breaks time-reversal symmetry while preserving inversion symmetry.
However, such magnetic order has eventually not been observed experimentally \cite{Hiraki, Fujiyama}, suggesting that our model in \cite{Ohki2020BETS} may have been oversimplified.
In fact, it has been known in Dirac electron systems that not only the onsite Hubbard interaction but also long-range repulsive interactions would play significant roles in determining the electronic properties \cite{Khveshchenko, Kotov, Hirata2021}.
Moreover, the treatment of SOC was incomplete in the sense that the coupling constant and the spin
vector were set as a constant, which is strictly speaking invalid in $\alpha$-(BETS)$_2$I$_3$ \cite{Winter}.
Therefore, it remains controversial as to how the repulsive interactions participate in the emergence of the insulating phase, and how they are related to the observed Berry phase change upon increasing $P$ \cite{Kawasugi}.

In the present study, to better understand the insulating phase of $\alpha$-(BETS)$_2$I$_3$ we will overcome these difficulties by performing more realistic microscopic calculations.
In particular, as shall be formulated in the succeeding parts, we will first employ ab-initio calculations to estimate realistic repulsive and spin-orbit interactions and then combine their results with an extended-Hubbard type model calculations to examine both impacts of SOC and next-nearest neighbor repulsions and discuss their relationship with possible topological states.

\subsection{Effective model based on first-principles calculation}
First-principles calculation was performed to derive an effective model using the X-ray crystal structural data of $\alpha$-(BETS)$_2$I$_3$ at 30K under ambient pressure \cite{KitouSawaTsumuraya}.
As the exchange-correlation function, GGA was used in the Quantum espresso (QE) package \cite{Perdew,Giannozzi}.
We used SG15 optimized norm-conserving Vanderbilt fully relativistic pseudopotentials to consider the effect of SOC in the first-principles calculation \cite{Schlipf}.
We set the cutoff energies of the wave functions and charge densities as 80 and 320 Ry, respectively, and the mesh of wave number ${\bm k}$ as $5\times5\times3$.
After the first-principles calculation, to obtain the transfer integrals with SOC, the maximally localized Wannier functions (MLWFs) were created using the Wannier90 code \cite{Wannier90}.
Eight bands near the Fermi energy were selected to construct the MLWFs, and the initial MLWF coordinates were set at the center of the BETS molecule in the unit cell.

\begin{figure}
\begin{centering}
\includegraphics[width=80mm]{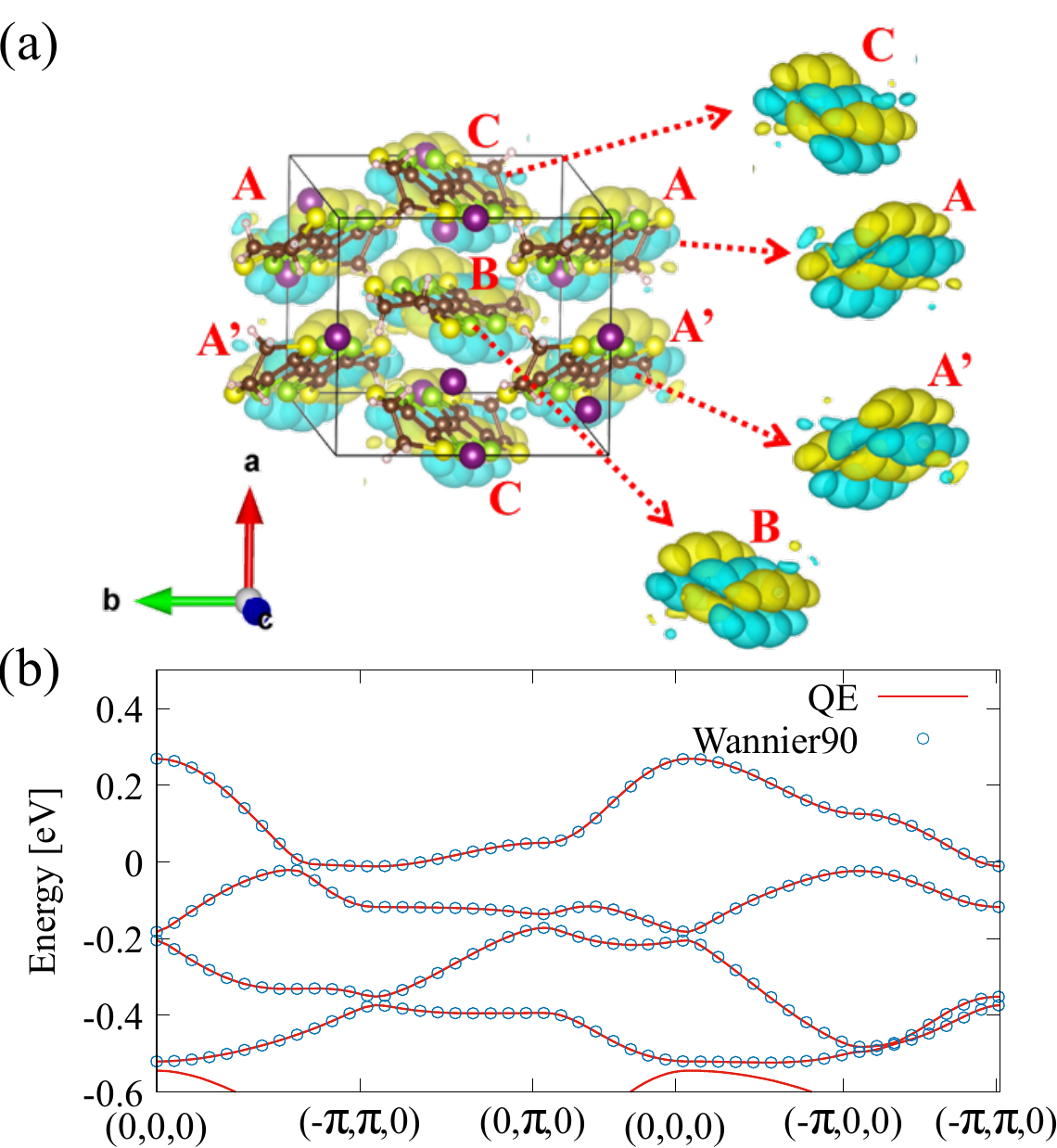}
\caption{(Color online) 
(a) Crystal structure and real-space distribution of the MLWFs for $\alpha$-(BETS)$_2$I$_3$ under ambient pressure at 30 K in a unit cell drawn by VESTA \cite{Momma}.
The box drawn by the black line represents the unit cell, and there are four BETS molecules (sites) labeled A, A', B, and C in the unit cell.
The A and A' sites are crystallographically equivalent but the A, B and C sites are non-equivalent.
(b) Energy band structure with SOC effect obtained by first-principles calculation.
}\label{Fig:Wannier-Fit}
\end{centering}
\end{figure}
Figure \ref{Fig:Wannier-Fit}(a) shows the crystal structure of the unit cell for $\alpha$-(BETS)$_2$I$_3$.
The box drawn by the black line represents the unit cell, and there are four BETS molecules (sites) labeled A, A', B, and C in the unit cell.
The A and A' sites are crystallographically equivalent but the A, B and C sites are non-equivalent.
The B and C sites, and the center of A and A' sites include inversion symmetric points.
The real-space distribution of the MLWFs at each unit cell site are also shown.
Figure \ref{Fig:Wannier-Fit}(b) shows the energy band structure with SOC calculated using QE and the Wannier interpolation {performed using} Wannier90.
The energy origin is set at the Fermi energy.

We also calculated the repulsive interactions considering the screening effect using the cRPA method in the RESPACK code \cite{Nakamura}.
We set the energy cutoff of the dielectric function as 5.0 Ry.

\begin{figure}
\begin{centering}
\includegraphics[width=85mm]{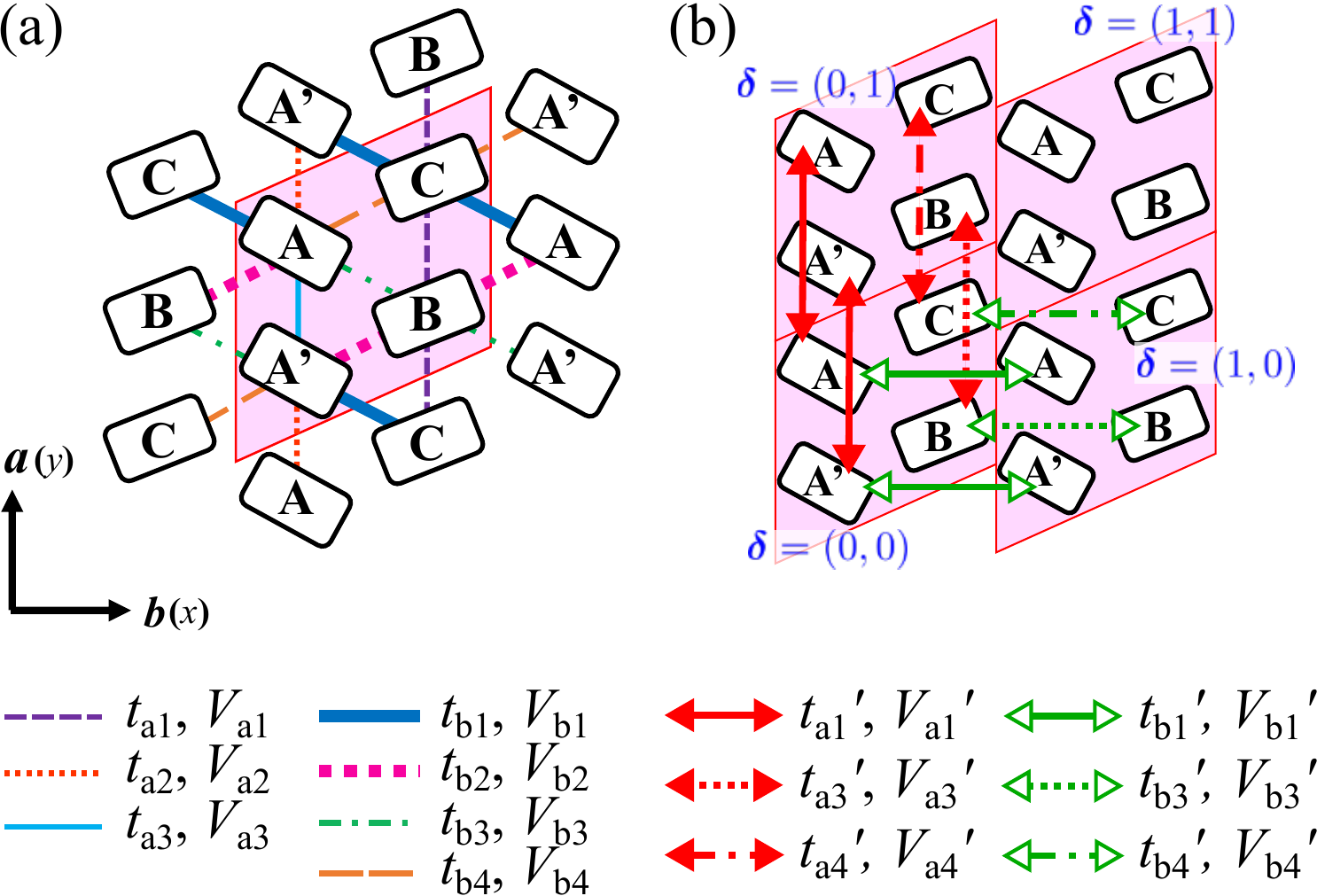}
\caption{(Color online) Schematic {of the} lattice structure of $\alpha$-(BETS)$_2$I$_3$ and definition of the transfer integrals and the repulsive interactions for the (a) nearest-neighbor sites, and (b) next-nearest-neighbor sites.
The shaded pink parallelogram {denotes a} the unit cell.
}\label{Fig:2Dnetwork}
\end{centering}
\end{figure}
Figure \ref{Fig:2Dnetwork} (a) and (b) show the two-dimensional (2D) lattice structures of $\alpha$-(BETS)$_2$I$_3$.
The nearest-neighbor and next-nearest-neighbor components of the transfer integrals and the interactions considered in our calculation are also illustrated.
In our calculation, spin polarization was not considered. 
However, as the energy scale of the spin polarization is small compared to the repulsive interactions, its effect on the repulsive interactions may not be significant.
The detailed values of transfer integrals and repulsive interactions are shown in Appendix A.

Based on the above first-principles calculation results, we constructed a 2D extended Hubbard model \cite{Seo2000}:
\begin{align}
H&=\sum_{\bm{R}, \bm{\delta}}\sum_{\alpha,\beta}\sum_{\sigma_1,\sigma_2}t^{({\bm \delta}){\rm SOC}}_{\alpha,\sigma_1; \beta,\sigma_2}c^{\dag}_{{\bm R},{\alpha},{\sigma_1}}c_{{\bm R}+\bm{\delta},\beta,{\sigma_2}}\nonumber\\
&+\xi{\sum_{{\bm R},{\alpha}}}U_{\alpha}n_{{\bm R},{\alpha},{\uparrow}}n_{{\bm R},{\alpha},{\downarrow}}\nonumber\\
&+\frac{\xi}{2}\sum_{{\bm R},{\bm \delta}}\sum_{\alpha,\beta}\sum_{\sigma_1,\sigma_2} V^{({\bm \delta})}_{\alpha,\beta}n_{{\bm R},\alpha,\sigma_1}n_{{\bm R}+{\bm \delta},\beta,\sigma_2},\label{Eq:Hamiltonian}
\end{align}
where ${\bm R}$ is the coordinate of the unit cell and ${\bm \delta}=(\delta_b,\delta_a)$ is the relative lattice vector in the $a$-$b$ plane.
$\alpha$ and $\beta$ are the site indices, and $\sigma_1$ and $\sigma_2$ are the spin indices ($\uparrow,\downarrow$).
$t^{({\bm \delta}){\rm SOC}}_{\alpha,\sigma_1; \beta,\sigma_2}$ is the transfer integral including SOC between $(\alpha,\sigma_1)$ and $(\beta,\sigma_2)$ separated by ${\bm \delta}$.
Site potentials, which is defined as the onsite components of transfer integrals ($t_{\alpha,\sigma_1;\alpha,\sigma_1}^{({\bm 0}){\rm SOC}}$), $t^{({\bm 0}){\rm SOC}}_{\rm A}=t^{({\bm 0}){\rm SOC}}_{\rm A'}=4.470$ eV, $t^{({\bm 0}){\rm SOC}}_{\rm B}=4.465$ eV, and $t^{({\bm 0}){\rm SOC}}_{\rm C}=4.477$ eV are excluded in our model because their contribution to the energy band is negligible.
$c^{\dag}_{{\bm R},{\alpha},{\sigma_1}}$ ($c_{{\bm R},{\alpha},{\sigma_1}}$) is the creation (annihilation) operator for the $\alpha$ site with spin $\sigma_1$ in the unit cell located at ${\bm R}$.
$U_\alpha$ and $V_{\alpha,\beta}^{({\bm \delta})}$ are the onsite and the inter-site repulsions, respectively, given by the static effective direct integrals $W_{\alpha,\beta}^{({\bm \delta})}$ calculated using RESPACK (see Appendix A).
Values of $W_{\alpha,\beta}^{({\bm \delta})}$ are a little too large to be used in mean-field calculation, although it is a value that considers the screening effect.
Therefore, in the following calculations, $W_{\alpha,\beta}^{({\bm \delta})}$ is multiplied by a constant $\xi$ ($0<\xi<1$) and the value is controlled.
We also defined the number operator as $n_{{\bm R},\alpha,\sigma}=c^\dag_{{\bm R},\alpha,\sigma}c_{{\bm R},\alpha,\sigma}$.
In the following, the lattice constants, Boltzmann constant $k_B$, and the Planck constant $\hbar$ are considered to be unity.
Furthermore, the electronvolt (eV) is used as the unit of energy throughout this paper, unless otherwise stated.

\subsection{Electronic state using Hartree-Fock approximation}

We treated Eq. (\ref{Eq:Hamiltonian}) within the Hartree-Fock approximation in the wave number space.
Fourier inverse transform, $c_{{\bm R},\alpha,\sigma_1}=N_{\rm cell}^{-1/2}\sum_{\bm k}c_{{\bm k},\alpha,\sigma_1}e^{i{\bm k}\cdot{\bm R}}$, was performed on Eq. (\ref{Eq:Hamiltonian}).
Here, $N_{\rm cell}$ is the total number of unit cells and ${\bm k}=(k_x, k_y)$ indicates the wave-number vector.
The Hartree-Fock Hamiltonian is as follows:
\begin{eqnarray}
\begin{aligned}
&H_{\rm HF}=H_T+H_U+H_V\\
&H_T=\sum_{\bm k}\sum_{\alpha,\beta}\sum_{\sigma_1,\sigma_2}\sum_{\bm \delta}t_{\alpha,\sigma_1;\beta,\sigma_2}^{({\bm \delta}){\rm SOC}}e^{i{\bm k}\cdot{\bm \delta}}c^\dag_{{\bm k},\alpha,\sigma_1}c_{{\bm k},\beta,\sigma_2}\\
&H_U=\xi \sum_{\bm k}\sum_{\alpha}U_\alpha\left[\sum_{\sigma_1}\langle n_{\alpha,\sigma_1}\rangle c^\dag_{{\bm k},\alpha,\bar{\sigma_1}}c_{{\bm k},\alpha,\bar{\sigma_1}}\right.\\
&\hspace{0.5cm}\left.-\frac{1}{N_{\rm cell}}\sum_{{\bm k'}, {\bm q}}\left(\langle c^\dag_{{\bm k}-{\bm q},\alpha,\uparrow} c_{{\bm k'},\alpha,\downarrow}\rangle c^\dag_{{\bm k'}+{\bm q},\alpha,\downarrow}c_{{\bm k},\alpha,\uparrow}\right.\right.\\
&\hspace{0.5cm}\left.\left.+\langle c^\dag_{{\bm k'}+{\bm q},\alpha,\downarrow} c_{{\bm k},\alpha,\uparrow}\rangle c^\dag_{{\bm k}-{\bm q},\alpha,\uparrow}c_{{\bm k'},\alpha,\downarrow}\right)\right]\\
&H_V=\frac{\xi}{2}\sum_{\bm k}\sum_{\alpha,\beta}\sum_{\sigma_1,\sigma_2}\sum_{\bm \delta}V_{\alpha,\beta}^{({\bm \delta})}\left[2\langle n_{\beta,\sigma_2} \rangle c^\dag_{{\bm k},\alpha,\sigma_1} c_{{\bm k},\alpha,\sigma_1}\right.\\
&\hspace{0.5cm}\left.-\frac{1}{N_{\rm cell}}\sum_{{\bm k'}, {\bm q}}e^{-i{\bm q}\cdot{\bm \delta}}\left( \langle c^\dag_{{\bm k}-{\bm q},\alpha,\sigma_1} c_{{\bm k'},\beta,\sigma_2}\rangle c^\dag_{{\bm k'}+{\bm q},\beta,\sigma_2}c_{{\bm k},\alpha,\sigma_1}\right.\right.\\
&\hspace{0.5cm}\left.\left. +\langle c^\dag_{{\bm k'}+{\bm q},\beta,\sigma_2} c_{{\bm k},\alpha,\sigma_1}\rangle c^\dag_{{\bm k}-{\bm q},\alpha,\sigma_1}c_{{\bm k'},\beta,\sigma_2} \right)\right],
\end{aligned}\label{Eq:original_HF_ham}
\end{eqnarray}
where $\bar{\sigma_1}=-\sigma_1$, and $\langle n_{\alpha,\sigma_1}\rangle = \langle c^\dag_{{\bm 0},\alpha,\sigma_1}c_{{\bm 0},\alpha,\sigma_1}\rangle$ is the charge density of site $\alpha$ and spin $\sigma_1$.
An order parameter in the Hartree-Fock approximation is obtained by calculating $\langle c^\dag c\rangle$.
When the calculation is performed using Eq. (\ref{Eq:original_HF_ham}), orders such as charge/spin density waves, bond order wave, and the interaction-induced QAH and QSH \tb{insulating phases} can happen with various periodicities.
We set ${\bm q}={\bm 0}$ and ${\bm k'}={\bm k}$ to exclude long-periodic orders, which have not been experimentally observed.
With these simplifications, the Fourier transform of the $\langle c^\dag c\rangle$ in the Hartree-Fock approximation is given by
\begin{eqnarray}
\begin{aligned}
&\langle c^\dag_{{\bm k},\alpha,\sigma_1}c_{{\bm k},\beta,\sigma_2} \rangle = \sum_{{\bm\delta}}\langle c^\dag_{{\bf 0},\alpha,\sigma_1}c_{{\bm \delta},\beta,\sigma_2} \rangle e^{-i{\bm k}\cdot{\bm \delta}},\\
&\langle c^\dag_{{\bm 0},\alpha,\sigma_1}c_{{\bm \delta},\beta,\sigma_2}\rangle = \frac{1}{N_{\rm cell}}\sum_{{\bm k},\nu,\sigma}\frac{d_{\alpha,\sigma_1;\nu,\sigma}({\bm k})d^*_{\beta,\sigma_2;\nu,\sigma}({\bm k})}{1+\exp{\left( E_{\nu,\sigma}({\bm k})/T\right)}}e^{i{\bm k}\cdot{\bm \delta}}.
\end{aligned}\nonumber\\
\end{eqnarray}
Thus, $H_{\rm HF}$ can finally be expressed as follows:
\begin{eqnarray}
\begin{aligned}
&H_{\rm HF}=\sum_{\bm k}\sum_{\alpha,\beta}\sum_{\sigma_1,\sigma_2}H_{\alpha,\sigma_1;\beta,\sigma_2}({\bm k})c^\dag_{{\bm k},\alpha,\sigma_1}c_{{\bm k},\beta,\sigma_2}\\
&H_{\alpha,\sigma_1;\beta,\sigma_2}({\bm k})=\sum_{{\bm \delta}} \Biggl{[}t_{\alpha,\sigma_1;\beta,\sigma_2}^{(\bm{\delta}){\rm SOC}}e^{i{\bm k}\cdot\bm{\delta}}\\
&\left.\hspace{0.5cm}-\xi V_{\alpha,\beta}^{({\bm\delta})} \left(\sum_{{\bm\delta'}} \langle c^\dag_{{\bm 0},\beta,\sigma_2}c_{{\bm \delta'},\alpha,\sigma_1}\rangle e^{-i{\bm k}\cdot{\bm \delta'}}\right) \right]\\
&\hspace{0.5cm}+\xi \delta_{\alpha,\beta}\biggl{[} U_\alpha\left( \langle n_{\alpha,\bar{\sigma_1}} \rangle\delta_{\sigma_1,\sigma_2}-\langle c^\dag_{{\bm 0},\alpha,\sigma_1} c_{{\bm 0},\alpha,\bar{\sigma_1}} \rangle \delta_{\bar{\sigma_1},\sigma_2} \right)\\
&\hspace{0.5cm}+\sum_{\gamma} \sum_{\sigma_3} \sum_{\bm \delta} V_{\alpha,\gamma}^{({\bm \delta})} \langle n_{\gamma,\sigma_3} \rangle \biggr{]},
\end{aligned}\label{Eq:HF_Hamiltonian}
\end{eqnarray}
where the off-diagonal site component $\langle c^\dag_{{\bm 0},\alpha,\sigma_1}c_{{\bm \delta},\beta,\sigma_2}\rangle$ ($\alpha\neq\beta$), which is included as the Fock term in Eq. (\ref{Eq:HF_Hamiltonian}), modulates the transfer integrals.
Order parameters in the interaction-induced QAH and QSH \tb{insulating phases} are defined as
\tb{
\begin{equation}
\chi^{\rm QAH({\bm \delta})}_{\alpha,\beta}=\sum_{\sigma_1}{\rm Im}\left[\langle c^\dag_{{\bm 0},\alpha,\sigma_1}c_{{\bm \delta},\beta,\sigma_1}\rangle\right],\label{Eq:QAH_op}
\end{equation}
for QAH \tb{insulating phase} and 
\begin{equation}
\chi^{\rm QSH({\bm \delta})}_{\alpha,\beta}=\sum_{\sigma_1}{\rm sgn}(\sigma_1){\rm Im}\left[\langle c^\dag_{{\bm 0},\alpha,\sigma_1}c_{{\bm \delta},\beta,\sigma_1}\rangle\right],\label{Eq:QSH_op}
\end{equation}
}
for QSH \tb{insulating phase}, respectively.
Here ${\rm sgn}(\sigma_1)=+1$ $(-1)$ for $\sigma_1=\uparrow$ ($\downarrow$).
\tb{
In the following, for simplicity, indices ($\alpha$, $\beta$, $\bm \delta$) are represented by the label $l={\rm a1}, {\rm a2}, \cdots, {\rm b4'}$ defined in Fig. \ref{Fig:2Dnetwork}.
}
$\chi^{\rm QAH}$ and $\chi^{\rm QSH}$ cause phase modulation in transfer integrals and open energy gap at the Dirac points \cite{Haldane, KaneMele, Raghu, Rachel, Omori2014}.

We numerically diagonalized $H_{\alpha,\sigma_1;\beta,\sigma_2}({\bm k})$ using the eigenvector $d_{\alpha,\sigma_1; \nu, \sigma}({\bm k})$ and obtained the energy eigenvalues.
For convenience, we define the eigenvalues $E_{\nu, \sigma}({\bm k})$ as
\begin{align}
&E_{\nu, \sigma}({\bm k})=\bar{E}_{\nu, \sigma}({\bm k})-\mu\nonumber\\
&=\biggl{\langle}\sum_{\alpha, \beta}\sum_{\sigma_1,\sigma_2} d^*_{\alpha,\sigma_1; \nu, \sigma}({\bm k}) H_{\alpha, \sigma_1;\beta, \sigma_2}({\bm k}) d_{\beta,\sigma_2; \nu, \sigma}({\bm k})\biggr{\rangle}-\mu,\nonumber\\
\end{align}
where $\mu$ is the chemical potential.
$\bar{E}_{\nu, \sigma}({\bm k})$ are the eigenvalues with band index $\nu$ obtained by numerical diagonalization [$\bar{E}_{1, \sigma}({\bm k})>\bar{E}_{2, \sigma}({\bm k})>\bar{E}_{3, \sigma}({\bm k})>\bar{E}_{4, \sigma}({\bm k})$].
In the calculation, we ignored spin orders such as a spin ordered massive Dirac electron phase suggested in our preceding study \cite{Ohki2020BETS} because the spin order is not consistent with the results in a recent NMR experiment \cite{Hiraki, Fujiyama}.
Calculation results when spin orders are allowed as a stable solution are shown in Appendices B and C.

Berry curvature, $B^z_{\nu,\sigma}({\bm k})$, and Chern number for spin $\sigma$, $N^{\rm Ch}_\sigma$, are defined as
\begin{eqnarray}
B^z_{\nu,\sigma}({\bm k}) &=& \sum_{\nu'\neq\nu}\frac{{\rm v}^x_{\nu,\nu',\sigma}({\bm k}){\rm v}^y_{\nu',\nu,\sigma}({\bm k})}{i\left[E_{\nu,\sigma}({\bm k})-E_{\nu',\sigma}({\bm k})\right]^2}+{\rm c.c.},\\
N^{\rm Ch}_\sigma &=& \frac{1}{2\pi}\int_{BZ}d{\bf k}B^z_{\nu,\sigma}({\bm k}),\label{Eq:chern}
\end{eqnarray}
where the velocity matrix ${\rm v}^\eta_{\nu,\nu',\sigma}({\bm k})$ along the $a$-axis ($\eta=a(y)$) and $b$-axis ($\eta=b(x)$) is given by
\begin{eqnarray}
{\rm v}^\eta_{\nu,\nu',\sigma}({\bm k}) &=& \sum_{\alpha,\beta}\sum_{\sigma_1,\sigma_2}d^*_{\alpha,\sigma_1;\nu,\sigma}({\bm k})\nonumber\\
&&\times\frac{\partial H_{\alpha,\sigma_1;\beta,\sigma_2}({\bm k})}{\partial k_\eta}d_{\beta,\sigma_2;\nu',\sigma}({\bm k}).
\end{eqnarray}
The Chern number $N^{\rm Ch}$ and spin Chern number $N^{\rm sCh}$ are calculated by
\begin{eqnarray}
N^{\rm Ch}&=&\sum_\sigma N^{\rm Ch}_\sigma,\label{Eq:chern}\\
N^{\rm sCh} &=& \sum_\sigma {\rm sgn}(\sigma)N^{\rm Ch}_\sigma.\label{Eq:spin_chern}
\end{eqnarray}
When the interaction-induced QAH and QSH \tb{insulating phases} occur, $N^{\rm Ch}$ and $N^{\rm sCh}$ become $\pm1$, and Hall and spin Hall conductivities $\sigma^{\rm H}$ and $\sigma^{\rm SH}$ are given by \cite{Haldane, KaneMele, Raghu, Rachel, Omori2014}
\begin{eqnarray}
\sigma^{\rm H}&=&-\frac{e^2}{h}N^{\rm Ch},\label{Eq:Hall}\\
\sigma^{\rm SH}&=&-\frac{e^2}{h}N^{\rm sCh}.\label{Eq:spinHall}
\end{eqnarray}

The Onsager phase factor $\gamma$ that characterizes quantum oscillations caused by quantization condition for the energy levels of electrons is also calculated to compare with results in Shubnikov-de Haas oscillation measurements \cite{Kawasugi}.
Based on the semiclassical theory \cite{Mikitik1999, Mikitik2012, Taskin, Wright2013, Georbig2011}, $\gamma$ is given by
\begin{eqnarray}
\begin{aligned}
&\gamma\equiv\frac{1}{2}-\frac{\phi_B}{2\pi},\\
&\phi_B=\int_{S_F}{B^z_{\nu,\sigma}({\bm k})}dS,\label{Eq:PhaseFactor}
\end{aligned}
\end{eqnarray}
where $\phi_B$ is the Berry phase and $\int_{S_F}$ means surface integration on the Fermi surface $S_F$.
In the Dirac electron system, $\Phi_B=\pi$ and $\gamma=0$ when degenerated point of Dirac cone is included in the integral range of surface integration for $\bm k$.
On the other hand, $\Phi_B=0$ and $\gamma=1/2$ in the other systems.
Therefore, we can confirm a change in the electronic state between the Dirac electron system and the normal electron system by looking at the $\gamma$ value switching.
$\gamma$ is calculated using an electron-doped band of approximately $0.005$ eV to reproduce the carrier doping performed in the recent experiment \cite{Kawasugi}.

\subsection{Conductivity}
The DC conductivity was calculated using the Nakano-Kubo formula \cite{Streda, Shon, Proskurin, Ruegg, Omori2017}.
The longitudinal DC conductivity along $\eta$-axis ($\eta=b(x), a(y)$) direction is expressed by
\begin{align}
&\sigma_\eta=\int d\omega \left(-\frac{df}{d\omega} \right)\Phi_\eta(\omega),\\
&\Phi_\eta(\omega)=\frac{2 e^2}{\pi N_{\rm cell}}\sum_{{\bm k}}\nonumber\\
&\times{\rm Tr}\left[ \hat{{\rm v}}^\eta({\bm k}) {\rm Im}\hat{G}^R({\bm k},\omega)\hat{{\rm v}}^\eta({\bm k}) {\rm Im}\hat{G}^R({\bm k},\omega) \right],\label{Eq:transport_function}
\end{align}
where $\hat{G}^R({\bm k},\omega)$ is the retarded Green's function, which is obtained by the analytic continuation of the Matsubara frequency $i\varepsilon_n$ ($\varepsilon_n = (2n+1)\pi T$) to a real frequency $\omega$.
We treated the effect of elastic scattering between electrons and the impurities originating from the lack and disorder of I$^{3-}$ molecules, using $T$-matrix approximation.
The impurity potential term $H_{\rm imp}$ is defined by
\begin{eqnarray}
H_{\rm imp}=\frac{V_0}{N_{\rm cell}}\sum_{{\bm k},{\bf q},\alpha,\sigma}\sum_{i}^{N_{\rm imp}}e^{-i{\bf q}\cdot {\bf r}_i}c^{\dag}_{{\bm k}+{\bf q},\alpha,\sigma}c_{{\bm k},\alpha,\sigma},
\end{eqnarray}
and we treated it using the perturbation theory for Green's function.
Here, $V_0 = 0.04$ is the strength of the potential and $\bm r_i$ ($i=1,\cdots,N_{\rm imp}$) is the coordinate of the $i$-th impurity.
We treated $H_{\rm imp}$ within $T$-matrix approximation, assuming that the impurity density $c_{\rm imp} = N_{\rm imp} / N_{\rm cell}$ is quite small ($c_{\rm imp} \ll 1$) and impurities are uniformly distributed.
We introduced the one body Green's function in the Hartree-Fock approximation calculated using $H_{\rm HF}$ (Eq. (\ref{Eq:HF_Hamiltonian})) as
\begin{eqnarray}
G^0_{\alpha,\sigma_1; \beta,\sigma_2}(\varepsilon_n, {\bm k}) = \sum_{\nu, \sigma} \frac{d_{\alpha,\sigma_1; \nu, \sigma}({\bm k}) d^*_{\beta,\sigma_2; \nu, \sigma}({\bm k})}{i\varepsilon_n - E_{\nu, \sigma}({\bm k})}.\label{Eq:one_body_g}
\end{eqnarray}
The retarded self-energy in $T$-matrix approximation $\Sigma^R_{\nu,\nu'}({\bm k},\omega)$ is calculated using the perturbation theory for Green's function.
When the real part of $\hat{G}^0$ and inter-band components of $\Sigma^R_{\nu,\nu'}({\bm k},\omega)$ ($\nu\neq\nu'$) can be ignored, the damping constant $\gamma_{\nu,\sigma}({\bm k},\omega)$ in the $T$-matrix approximation is obtained as follows:
\begin{eqnarray}
\gamma_{\nu,\sigma}({\bm k},\omega)&=&\frac{\hbar}{2\tau_{\nu,\sigma}({\bm k},\omega)}=-{\rm Im}\Sigma^R_{\nu,\sigma}({\bm k},\omega)\nonumber\\
&=&c_{\rm imp}\frac{ \left|d_{\alpha,\nu,\sigma}({\bm k})\right|^2\left\{ \pi V_0^2 \rho^E_\sigma(\omega)\right\}}{1+\left\{ \pi V_0 \rho^E_\sigma(\omega)\right\}^2},
\end{eqnarray}
where $c_{\rm imp}=\frac{N_{\rm imp}}{N_{\rm cell}}=0.02 \ll 1$ is the impurity density and $\tau_{\nu,\sigma}({\bm k},\omega)$ is the relaxation time in $T$-matrix approximation.
$\rho^E_\sigma(\omega)$ represents the density of states for $\sigma$:
\begin{align}
\rho^E_\sigma(\omega) = \frac{1}{N_{\rm cell}}\sum_{{\bm k},\alpha,\sigma_1,\nu}|d_{\alpha,\sigma_1; \nu, \sigma}({\bm k})|^2\delta(\hbar\omega-E_{\nu,\sigma}({\bm k})).
\end{align}
We ignored the real part of $\Sigma^R_{\nu}({\bm k},\omega)$ because it only gives the constant shift of energy in the limit of $c_{\rm imp}\ll1$ \cite{Omori2017}.
In this case, $\Phi_\eta(\omega)$ in Eq. (\ref{Eq:transport_function}) can be represented as follows:
\begin{align}
\Phi_\eta(\omega)=\frac{2 e^2}{N_{\rm cell}}\sum_{{\bm k},\nu,\sigma}\left|{\rm v}^\eta_{\nu,\sigma}({\bm k})\right|^2\tau_{\nu,\sigma}({\bm k},\omega)\delta(\hbar \omega-E_{\nu,\sigma}({\bm k})).
\end{align}
In this study, the DC conductivity is normalized to the universal conductivity $\sigma_0=4e^2/\pi h$.

\subsection{Calculation of the spin fluctuations using RPA}
We investigated the effect of spin fluctuations on the NMR properties, e.g., the Knight shift and $1/T_1T$ in the high-$T$ Dirac electron phase in RPA using eq. (\ref{Eq:Hamiltonian}) \cite{Kobayashi2004, Kobayashi2013, Matsuno2017}.

In the linear response theory, the irreducible susceptibility $\chi_{\alpha,\beta}^0({\bm q},\omega_m)$ can be calculated using the one body Green's function $\hat{G}^0$ (Eq. (\ref{Eq:one_body_g})) as follows:
\begin{align}
&\chi^0_{\alpha,\sigma_1;\beta,\sigma_2}({\bm q},\omega_m)\nonumber\\
&=-\frac{T}{N_{\rm cell}}\sum_{{\bm k},n}G^0_{\alpha,\sigma_1,\beta,\sigma_1}({\bm k}+{\bm q},\varepsilon_n+\omega_m)G^0_{\beta,\sigma_2,\alpha,\sigma_2}({\bm k},\varepsilon_n)\nonumber\\
&=\frac{1}{N_{\rm cell}}\sum_{\bm k}\sum_{\sigma,\sigma'}\sum_{\nu,\nu'=1}^4F_{\alpha,\sigma_1;\beta,\sigma_2}^{\nu,\sigma;\nu',\sigma'}({\bm k},{\bm q})\chi^{0}_{\nu,\sigma; \nu',\sigma'}({\bm q},\omega_m),\\
&\chi^{0}_{\nu,\sigma;\nu',\sigma'}({\bm q},\omega_m)=-\frac{f(E_{\nu,\sigma}({\bm k}+{\bm q}))-f(E_{\nu',\sigma'}({\bm k}))}{E_{\nu,\sigma} ({\bm k}+{\bm q})-E_{\nu',\sigma'}({\bm k})+i\omega_m},
\end{align}
where $N_{\rm cell}$ is the system size and $\omega_m = 2m\pi T$.
$f(E_{\nu,\sigma}({\bm k}))=\left[ 1 + \exp(E_{\nu,\sigma}({\bm k})/T) \right]^{-1}$ represents the Fermi distribution function.
$F({\bm k},{\bm q})$ indicates the form factor represented by
\begin{eqnarray}
F_{\alpha,\sigma_1;\beta,\sigma_2}^{\nu,\sigma;\nu',\sigma'}({\bm k},{\bm q}) &=& d_{\alpha,\sigma_1,\nu,\sigma}({\bm k}+{\bm q})d^*_{\beta,\sigma_1;\nu,\sigma}({\bm k}+{\bm q}) \nonumber\\
&&\times d_{\beta,\sigma_2;\nu',\sigma'}({\bm k})d^*_{\alpha,\sigma_2;\nu',\sigma'}({\bm k}).
\end{eqnarray}

In RPA, the spin susceptibility $\hat{\chi}^S$ and the transverse spin susceptibility $\hat{\chi}^\pm$ in the absence of an external field and {in} presence of spin symmetry are calculated as follows.
\begin{eqnarray}
\hat{\chi}^S({\bm q},\omega) &=& \hat{\chi}^\pm({\bm q},\omega)\nonumber\\
&=& \left( \hat{\bm I}-\hat{\chi}^0({\bm q},\omega)\hat{U}\right)^{-1}\hat{\chi}^0({\bm q},\omega),
\end{eqnarray}
where $\hat{\bm I}$ is the unit matrix and $\hat{U} = \xi U_\alpha\delta_{\alpha, \beta}$.
Moreover, to estimate the contribution of the intra- and inter-band components to the spin fluctuations, we divided the irreducible susceptibility into two components \cite{Matsuno2017}:
\begin{eqnarray}
\chi^{0,{\rm Intra}}_{\alpha,\sigma_1;\beta,\sigma_2}({\bm q},\omega)&=&\frac{1}{N_{\rm cell}}\sum_{\bm k}\sum_{\sigma,\sigma'}\sum_{\nu=\nu'}F_{\alpha,\sigma_1;\beta,\sigma_2}^{\nu,\sigma;\nu',\sigma'}({\bm k},{\bm q})\nonumber\\
&&\times\chi^{0}_{\nu,\sigma;\nu',\sigma'}({\bm q},\omega),\\
\chi^{0,{\rm Inter}}_{\alpha,\sigma_1;\beta,\sigma_2}({\bm q},\omega)&=&\frac{1}{N_{\rm cell}}\sum_{\bm k}\sum_{\sigma,\sigma'}\sum_{\nu\neq\nu'}F_{\alpha,\sigma_1;\beta,\sigma_2}^{\nu,\sigma;\nu',\sigma'}({\bm k},{\bm q})\nonumber\\
&&\times\chi^{0}_{\nu,\sigma;\nu',\sigma'}({\bm q},\omega),
\end{eqnarray}
where $\hat{\chi}^{0,{\rm Intra}}$ and $\hat{\chi}^{0,{\rm Inter}}$ are the intra- and inter-band components of the irreducible susceptibility $\hat{\chi}^0$.
Thus, the intra-band component of the spin susceptibility is calculated as follows:
\begin{eqnarray}
\hat{\chi}^{S,{\rm Intra}}&=&\left( \hat{\bm I}-\hat{\chi}^{0,{\rm Intra}}\hat{U}\right)^{-1}\hat{\chi}^{0,{\rm Intra}}.\label{eq:chi_s_intra}
\end{eqnarray}
The inter-band component $\hat{\chi}^{S,{\rm Inter}}$ is also obtained based on the definition $\hat{\chi}^{S,{\rm Inter}}=\hat{\chi}^S-\hat{\chi}^{S,{\rm Intra}}$. 
Applying the analytical continuation $i\omega_m \to \hbar\omega + i0^+$, the site-resolved Knight shift $K_\alpha$ and $1/T_1T$ in RPA are obtained as follows:
\begin{eqnarray}
K_\alpha = \sum_\beta {\rm Re}\left[\chi^S_{\alpha,\beta}({\bm q}={\bm 0},\omega=0)\right],
\label{eq:knight_shift}
\end{eqnarray}
and
\begin{eqnarray}
1/T_1T = \sum_{\bm q}\sum_{\alpha}\frac{{\rm Im}\left[\chi^\pm_{\alpha,\alpha}({\bm q},\omega_0)\right]}{\omega_0},
\end{eqnarray}
where the frequency $\omega_0$ is infinitely close to zero and is set as $\omega_0 = 0.001$ in this study.

%
\section{Numerical Results}
%
\subsection{Electronic state at low-$T$ when a static effective direct integral is used}
In this subsection, to investigate more realistic orders which are possible caused in $\alpha$-(BETS)$_2$I$_3$ at low-$T$ region, the numerical results using the transfer integrals with SOC $t_{\alpha,\sigma_1;\beta,\sigma_2}^{({\bm \delta}){\rm SOC}}$ and the static effective direct integral $W_{\alpha,\beta}^{({\bm \delta})}$ obtained based on the first-principles calculation are shown.
We set the initial states of the mean-field calculation randomly and investigated the electronic state with the lowest energy, other than the charge, spin, and bond orders.
Throughout this subsection, $T = 1\times10^{-4}$ and we sweep the control parameter $\xi$ which is multiplied to $W_{\alpha,\beta}^{({\bm \delta})}$ from 0 (non-interacting case) to 1 to investigate the effect of the repulsive interaction to the electronic state.
As a result, we found that the interaction-induced QSH \tb{insulating phase} proposed in previous studies for the honeycomb lattice model \cite{Raghu, Rachel, Weeks, Dauphin} is the most stable \tb{solution} at low-$T$ region in $\alpha$-(BETS)$_2$I$_3$.

%
\begin{figure}
\begin{centering}
\includegraphics[width=85mm]{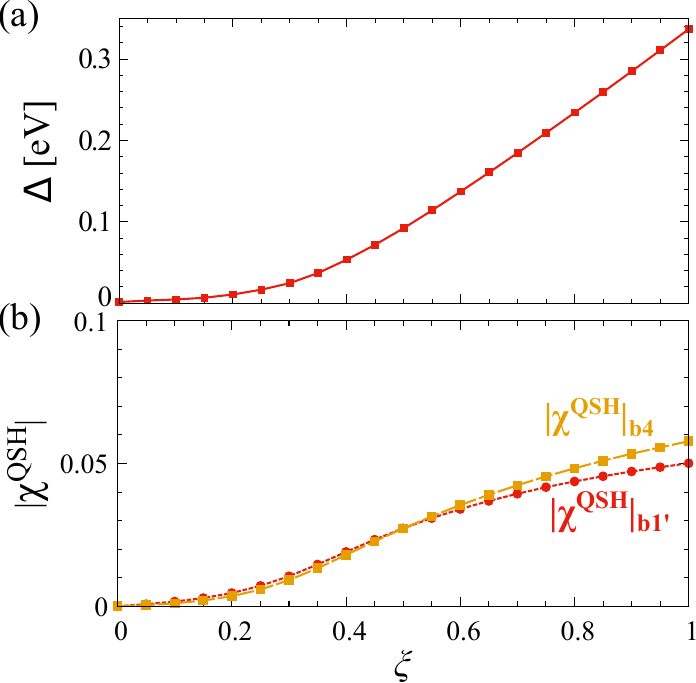}
\caption{(Color online)
$\xi$-dependence of the (a) energy gap $\Delta$, and (b) \tb{the absolute values of order parameter in the interaction-induced QSH insulating phase} which modulate transfer integrals $t_{b1}'$ and $t_{b4}$: $|\chi^{\rm QSH}|_{b1'}$ and $|\chi^{\rm QSH}|_{b4}$, respectively.
}\label{Fig:lambda-dep}
\end{centering}
\end{figure}
Figure \ref{Fig:lambda-dep}(a) shows the $\xi$-dependence of the energy gap at the Dirac point $\Delta$.
In $\xi=0$ (non-interacting case), the electronic system becomes a topological insulator (TI) and a small energy gap $\Delta\simeq0.002$ opens owing to the contribution of only the SOC term in the transfer integrals $t_{\alpha,\sigma_1;\beta,\sigma_2}^{({\bm \delta}){\rm SOC}}$.
With an increase in $\xi$, $\Delta$ increases continuously by the contribution of the interaction $W_{\alpha,\beta}^{({\bm \delta})}$.
In Fig. \ref{Fig:lambda-dep}(b), the $\xi$-dependence of \tb{the absolute values of order parameter $|\chi^{\rm QSH}|_{b1'}$ and $|\chi^{\rm QSH}|_{b4}$ defined by Eq. (\ref{Eq:QSH_op})} are plotted.
\tb{$\chi^{\rm QSH}_{b1'}$ ($|\chi^{\rm QSH}|_{b4}$)} contributes to the phase modulation of the next-nearest-neighbor (nearest-neighbor) component of the transfer integral $t_{b1}'$ ($t_{b4}$) illustrated in Fig. \ref{Fig:2Dnetwork}.
The signs of them are inverted according to the degrees-of-freedom of the spin $\sigma$ under the interaction-induced QSH insulating phase, as plotted in Fig. \ref{Fig:lambda-dep}(b).
In $\xi>0$, \tb{the order parameter} such as \tb{$\chi^{\rm QSH}_{b1'}$ and $\chi^{\rm QSH}_{b4}$} which is the same as the interaction-induced QSH \tb{insulating phase} proposed in previous studies for the honeycomb lattice model \cite{Raghu, Rachel, Weeks, Dauphin} becomes finite and is enhanced with the increase in $\xi$.
On the other hand, in the TI state ($\xi=0$), \tb{$\chi^{\rm QSH}_{b1'}$ and $\chi^{\rm QSH}_{b4}$} are zero.

%
\begin{figure}
\begin{centering}
\includegraphics[width=85mm]{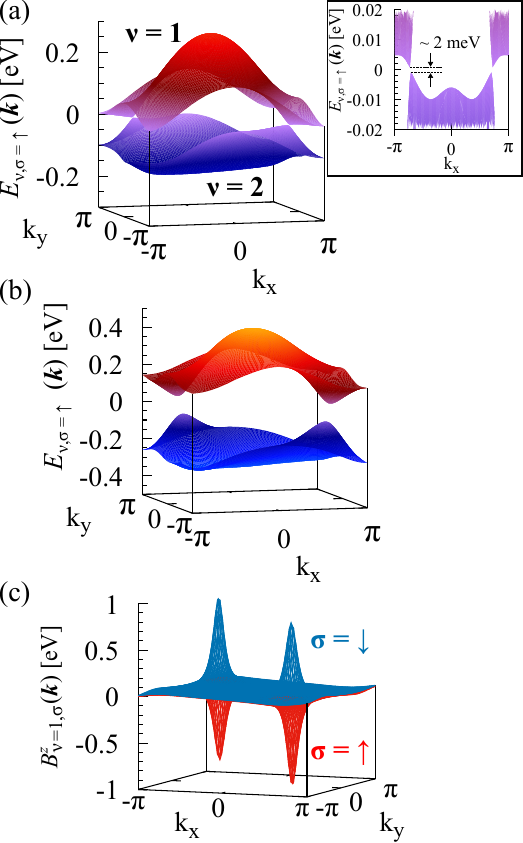}
\caption{(Color online)
The energy band structure $E_{\nu,\sigma}({\bm k})$ in the (a) TI ($\xi=0$) and (b) interaction-induced QSH insulator ($\xi=0.5$) states near the Fermi energy ($\nu = 1, 2$) for $\sigma = \uparrow$.
$E_{\nu,\uparrow}({\bm k})$ and $E_{\nu,\downarrow}({\bm k})$ are degenerated.
The inset in (a) is the magnified view of $E_{\nu,\sigma}({\bm k})$ in the TI state in the $-0.02<E_{\nu,\sigma}({\bm k})<0.02$ energy range.
(c) The Berry curvature $B^z_{\nu,\sigma}({\bm k})$ in the interaction-induced QSH \tb{insulating phase} for $\nu = 1$.
}\label{Fig:band-berrycurv}
\end{centering}
\end{figure}
%
Figures \ref{Fig:band-berrycurv}(a) and (b) show the energy eigenvalues $E_{\nu, \sigma=\uparrow}({\bm k})$ near the Fermi energy ($\nu=1,2$) in the TI ($\xi=0$) and the interaction-induced QSH insulator ($\xi=0.5$) states, respectively.
The energy gap in the TI state is approximately 2 meV as shown in the inset of Fig. \ref{Fig:band-berrycurv}(a).
The Berry curvature $B^z_{\nu,\sigma}({\bm k})$ under the interaction-induced QSH \tb{insulating phase} ($\xi=0.5$) is plotted in Figure \ref{Fig:band-berrycurv} (c) for each spin.
$B^z_{\nu,\sigma}({\bm k})$ has two peaks originating from two massive Dirac cones in the Brillouin zone.
The spin Chern number $N^{\rm sCh}$ defined in Eq. (\ref{Eq:spin_chern}) becomes $-1$ under the interaction-induced QSH \tb{insulating phase} because the two peaks of $B^z_{\nu,\sigma}({\bm k})$ have the same sign, which invert according to the spin index $\sigma$.
Therefore, the spin Hall conductivity $\sigma^{\rm SH}$ calculated by Eq. (\ref{Eq:spinHall}) is given by $\sigma^{\rm SH}=e^2/h$.
These wavenumber and spin dependencies on $B^z_{\nu,\sigma}({\bm k})$ are approximately the same as those in the interaction-induced QSH \tb{insulating phase} of the honeycomb lattice model \cite{Raghu, Rachel} and TI state of $\alpha$-(BETS)$_2$I$_3$ \cite{Ohki2020BETS}.
Therefore, $N^{\rm sCh}=\pm1$ and $\sigma^{\rm SH}=\mp e^2/h$ are obtained in both the TI and interaction-induced QSH \tb{insulating phases}.
In other words, the edge spin current in both states is quantized by the spin Chern number $N^{\rm sCh}$ owing to the bulk-edge correspondence.
The difference between the TI state \cite{TsumurayaSuzumura, SuzumuraTsumuraya, Winter} and the interaction-induced QSH \tb{insulating phase} \cite{Raghu, Rachel} is the energy gap values and order parameters:
The TI state has a slight gap of approximately 2 meV owing to the SOC contribution alone \cite{TsumurayaSuzumura, SuzumuraTsumuraya, Ohki2020BETS} and is not an ordered state because any corresponding order parameter, $\langle c^\dag c\rangle$, does not exist.
Conversely, in the interaction-induced QSH \tb{insulating phase}, the energy gap value depends on the magnitude of \tb{the order parameter} $\chi^{\rm QSH}$ defined by Eq. (\ref{Eq:QSH_op}), as shown in Figs. \ref{Fig:lambda-dep} (a) and (b).
The TI state is a band insulator caused by the contribution of intrinsic SOC alone, whereas the interaction-induced QSH \tb{insulating phase} is an ordered state derived from spontaneous symmetry breaking caused by the contribution of the repulsive interaction.

\subsection{Stability of the interaction-induced QSH \tb{insulating phase} in the presence of the repulsive interaction}
Next, to investigate the relationship between the stability of the interaction-induced QSH \tb{insulating phase} at low-$T$ in $\alpha$-(BETS)$_2$I$_3$ and the values of the nearest-neighbor and next-nearest-neighbor repulsions, we represent the nearest-neighbor and next-nearest-neighbor repulsions by parameters $V$ and $V'$, respectively, and draw the phase diagram for these parameters.
We also calculated the phase modulation in the interaction-induced QSH \tb{insulating phase} 
\begin{eqnarray}
\varphi=\tan^{-1}\frac{{\rm Im}[\langle c^\dag c\rangle]}{{\rm Re} [\langle c^\dag c\rangle]},\label{Eq:phase}
\end{eqnarray}
for several closed loops in the unit cell and investigate the effect of a local magnetic flux owing to the interaction-induced QSH insulating phase.
In this section, we show the calculation results for the following two cases to investigate the effect of the SOC term in transfer integrals: (1) when transfer integrals with SOC, $t_{\alpha,\beta,\sigma}^{({\bm \delta}){\rm SOC}}$, are used, and (2) when transfer integrals without SOC, $t_{\alpha,\beta,\sigma}^{({\bm \delta})}$, are used (see Table \ref{tab:TransWeff} in Appendix A).
Throughout this subsection, the onsite repulsion $U$ and $T$ are fixed at $(U, T) = (0.5, 1\times10^{-4})$, unless otherwise stated.

\subsubsection{$V$-$V'$ phase diagram when SOC exists}
\begin{figure}
\begin{centering}
\includegraphics[width=80mm]{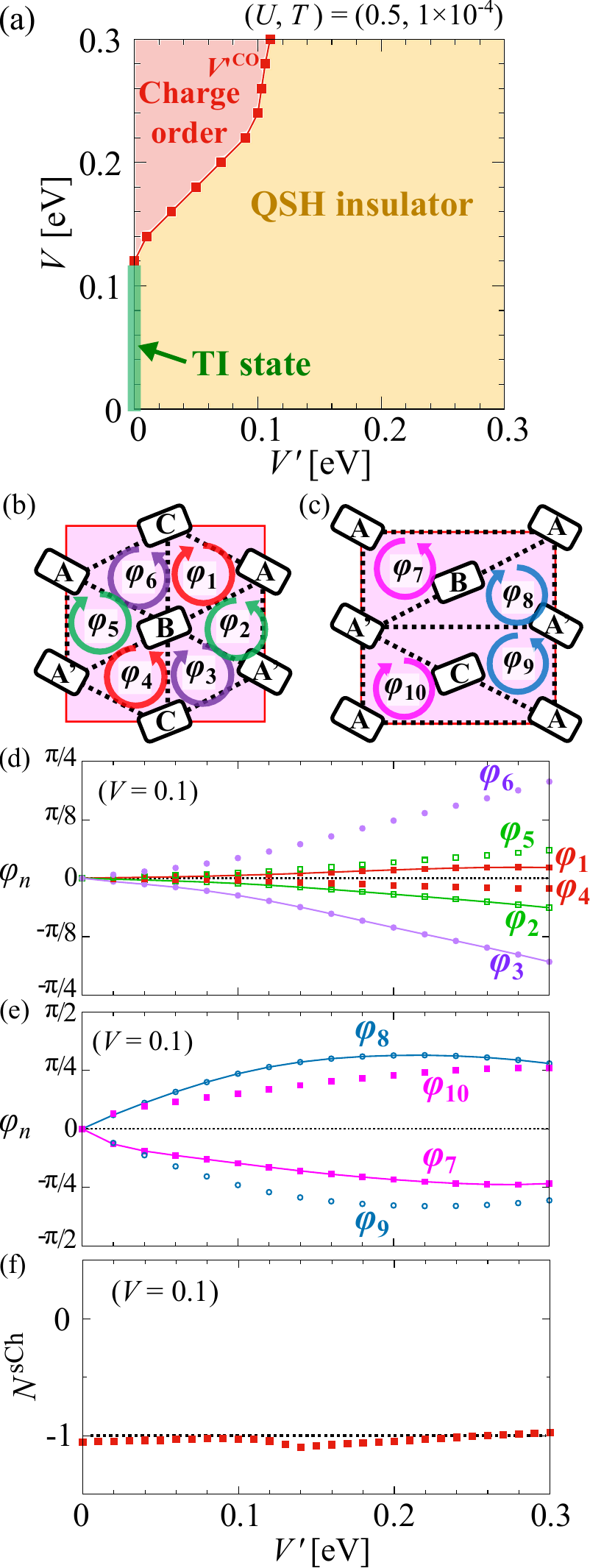}
\caption{(Color online)
(a) $V$-$V'$ phase diagram with SOC for $(U, T) = (0.5, 1\times10^{-4})$.
For $V' < V'^{\rm CO}$, the inversion symmetry is broken ($\langle n_{\rm A} \rangle \neq \langle n_{\rm A'} \rangle$) and the horizontal stripe charge ordered insulating phase appears.
(b) and (c) show the schematic of the unit cell and definition of the phases around the (b) three nearest-neighbor bonds and (c) three nearest-neighbor bonds and one next-nearest-neighbor bond.
(d) and (e) represent the $V'$-dependence of the phases on the loops shown in (b) and (c): $\varphi_n$ ($n = 1,...,10$) at $(U,V,T) = (0.5, 0.1, 1\times10^{-4})$ as an example.
\tb{
(f) $V'$-dependence of spin Chern number $N^{\rm sCh}$ at $(U,V,T) = (0.5, 0.1, 1\times10^{-4})$.
Dotted line of integer $-1$ is drawn for eye guide.
}
}\label{Fig:PDG_with_SOC}
\end{centering}
\end{figure}
We first draw the $V$-$V'$ phase diagram with SOC.
The calculation result is shown in Fig. \ref{Fig:PDG_with_SOC} (a).
When SOC is considered, the interaction-induced QSH \tb{insulating phase} is stabilized at $V'>0$ and $V<0.12$.
This indicates that $V'$ plays a significant role in stabilizing the interaction-induced QSH \tb{insulating phase} in $\alpha$-(BETS)$_2$I$_3$, as already suggested in studies for the honeycomb lattice model \cite{Raghu, Rachel}.
As indicated in the list of static effective direct integrals evaluated by RESPACK \cite{Nakamura} in Table \ref{tab:TransWeff}, $V'/V$ is expected to be large in $\alpha$-(BETS)$_2$I$_3$.
This tendency favors the realization of the interaction-induced QSH insulating phase.
The horizontal stripe charge ordered insulating phase \cite{Seo2000} appears in $V>0.12$ and $V'<V'^{\rm CO}$ (upper-left region).
However, the charge ordered insulating phase is not realistic in $\alpha$-(BETS)$_2$I$_3$ because inversion symmetry breaking has not been observed by the synchrotron X-ray diffraction experiment \cite{KitouSawaTsumuraya}.
The TI state caused only by the SOC contribution appears at $V'=0$ and $V<0.12$.

Next, we set $V=0.1$ and calculate the amount of phase modulation caused by the interaction-induced QSH \tb{insulating phase} $\varphi$ defined as Eq. (\ref{Eq:phase}) in the unit cell and investigated the presence or absence of a local magnetic flux.
Figures \ref{Fig:PDG_with_SOC} (b) and (c) show the schematic of the unit cell of $\alpha$-(BETS)$_2$I$_3$ and the loop patterns to calculate the summation of $\varphi$.
These loops include only the nearest-neighbor and next-nearest-neighbor bonds.
The $V'$-dependence of the summation of phases in each loop of the unit cell $\varphi_n$ ($n=1,\cdots,10$) at $V = 0.1$ are plotted in Figs. \ref{Fig:PDG_with_SOC} (d) and (e).
Signs of $\varphi_n$ depend on the spin degrees-of-freedom $\sigma$ under the interaction-induced QSH insulating phase.
The sum of $\varphi_n$ in the unit cell becomes zero because cancellation occurs: $\varphi_1=-\varphi_4$, $\varphi_2=-\varphi_5$, $\varphi_3=-\varphi_6$ (loops including only nearest-neighbor bonds) and $\varphi_7=-\varphi_{10}$, $\varphi_8=-\varphi_9$ (loops including nearest-neighbor and next-nearest-neighbor bonds).
Therefore, no total magnetic flux in the unit cell exists under the interaction-induced QSH insulating phase.
$\varphi_n$ is zero at only $V'=0$ and continuously increased as $V'$ is increased.

\tb{
Figure \ref{Fig:PDG_with_SOC} (f) shows the $V'$-dependence of spin Chern number $N^{\rm sCh}$ at $(U,V,T) = (0.5, 0.1, 1\times10^{-4})$.
Dotted line of integer $-1$ is also drawn.
$N^{\rm sCh}$ is almost $-1$ in both interaction-induced QSH insulating phase and TI state.
Note that the deviation of $N^{\rm sCh}$ from $-1$ is due to errors in numerical integration in discrete wavenumber space near the Dirac points.
}

\subsubsection{$V$-$V'$ phase diagram when SOC is absence}
\begin{figure}
\begin{centering}
\includegraphics[width=85mm]{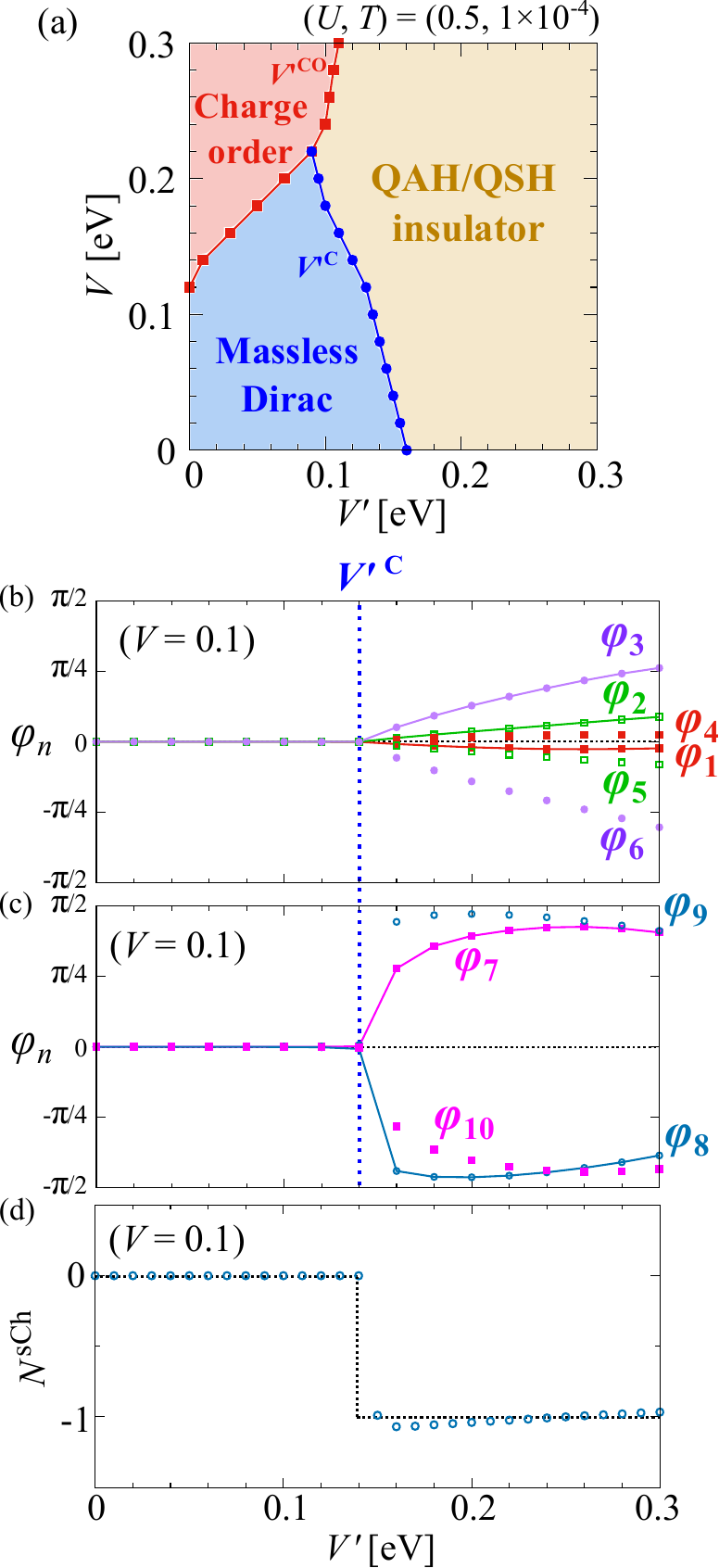}
\caption{(Color online)
(a) $V$-$V'$ phase diagram without SOC for $(U, T) = (0.5, 1\times10^{-4})$.
$V'^{\rm C}$ indicates the value of $V'$ that opens the energy gap.
(b) and (c) represent the $V'$-dependence of the phases on the loops shown in Figs. \ref{Fig:PDG_without_SOC} (b) and (c): $\varphi_n$ ($n = 1,...,10$) at $(U,V,T) = (0.5, 0.1, 1\times10^{-4})$ as an example.
\tb{
(d) $V'$-dependence of spin Chern number $N^{\rm sCh}$ at $(U,V,T) = (0.5, 0.1, 1\times10^{-4})$.
Dotted lines of integers zero and $-1$ are drawn for eye guide.
}
}\label{Fig:PDG_without_SOC}
\end{centering}
\end{figure}
%
Next, we used transfer integrals without the SOC term, $t_{\alpha,\beta,\sigma}^{({\bm \delta})}$, in the calculation and draw the $V$-$V'$ phase diagram without SOC.
The calculation result is shown in Fig. \ref{Fig:PDG_without_SOC} (a).
When SOC is absent, both the interaction-induced QSH and quantum anomalous Hall (QAH) \tb{insulating phases} can appear at $V'>V'^C$ and $V'>V'^{\rm CO}$ and these states are degenerated.
In the interaction-induced QAH insulating phase, ${\rm }[\langle c^\dag_\sigma c_\sigma\rangle]$ does not depend on the degrees-of-freedom of the spin $\sigma$.
\tb{The order parameter} $\chi^{\rm QAH}$ defined by Eq. (\ref{Eq:QAH_op}) has a finite value and the Chern number $N^{\rm Ch}$ defined in Eq. (\ref{Eq:chern}) becomes $\pm1$ and Hall conductivity defined in Eq. (\ref{Eq:Hall}) is given by $\sigma^{\rm H}=\mp{e^2}{h}$.
With the decrease in $V$ and $V'$, the massless Dirac electron phase appears owing to phase transition at $V'<V'^{\rm CO}$ and $V'<V'^{\rm C}$.

The $V'$-dependence of the summation of phases on several loops as shown Figs. \ref{Fig:PDG_without_SOC} (b) and (c), $\varphi_n$ ($n=1,\cdots,10$), at $V = 0.1$ are plotted in Figs. \ref{Fig:PDG_without_SOC} (b) and (c).
The sum of $\varphi_n$ in the unit cell cancelled each other out: $\varphi_1=-\varphi_4$, $\varphi_2=-\varphi_5$, $\varphi_3=-\varphi_6$ (loops including only the nearest-neighbor bonds) and $\varphi_7=-\varphi_{10}$, $\varphi_8=-\varphi_9$ (loops including the nearest-neighbor and next-nearest-neighbor bonds).
$\varphi_n$ vanishes suddenly at $V'=V'^C$ because the QAH \tb{insulating phase} to the massless Dirac electron phase transition occurs.

\tb{
Figure \ref{Fig:PDG_without_SOC} (d) shows the $V'$-dependence of spin Chern number $N^{\rm sCh}$ at $(U,V,T) = (0.5, 0.1, 1\times10^{-4})$.
Dotted lines of integers zero and $-1$ are also drawn.
$N^{\rm sCh}$ is almost $-1$ in the interaction-induced QSH insulating phase and quickly changes to zero when massless Dirac electron phase appears.
}

It can be considered that the contribution of SOC cannot be ignored in real material, so the QAH to massless Dirac electron phase transition does not occur in $\alpha$-(BETS)$_2$I$_3$.
The difference of the interaction-induced QSH \tb{insulating phase} with SOC and QAH \tb{insulating phase} without SOC in our results is analogous to a ferromagnet with and without an external magnetic field.
When SOC is considered, as presented in the previous subsection, the interaction-induced QSH \tb{insulating phase} at $V'>0$ continuously changes to the TI state at $V'=0$ with a decrease in $V'$.

\subsection{Switching of Onsager phase factor by contribution of the interaction}
In this subsection, we calculated the Onsager phase factor $\gamma$ under the interaction-induced QSH \tb{insulating phase} using Eq. (\ref{Eq:PhaseFactor}) and discuss the consistency with experimental results \cite{Kawasugi}.
It has been reported by Shubnikov-de Haas oscillation measurements that the Onsager phase factor $\gamma$ clearly switches $1/2$ to zero at the hydrostatic pressure $P=0.5$GPa at which the insulating phase of $\alpha$-(BETS)$_2$I$_3$ vanishes in the entire $T$-region \cite{Kawasugi}.
This result suggests that electron correlation effects cannot be negligible in $\alpha$-(BETS)$_2$I$_3$ similar to $\alpha$-(ET)$_2$I$_3$ under hydrostatic pressure \cite{TanakaOgata, Ishikawa, Beyer, Liu, Ohki2019,Kobayashi2013, HirataNat2016, Matsuno2017, Matsuno2018, HirataScience, Ohki2020}.
In this subsection, we considered the $P$-dependence as the change in $V'$ associated with the change in $P$ and calculate $V'$-dependence of $\gamma$ at low-$T$ to explain the experimental results.
Throughout this subsection, $T = 1\times10^{-4}$, $U=0.5$, and $V=0.1$ to investigate the contribution of the interaction-induced QSH \tb{insulating phase} to $\gamma$.

\begin{figure}
\begin{centering}
\includegraphics[width=75mm]{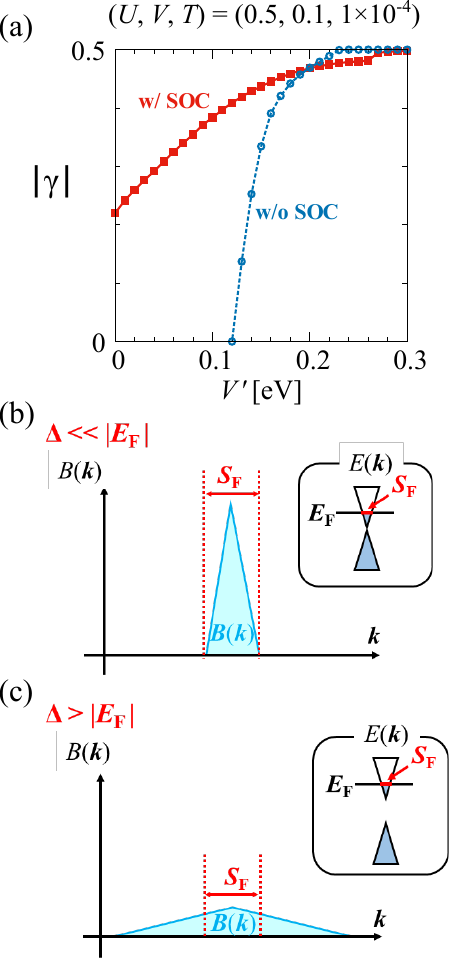}
\caption{(Color online)
(a) $V'$-dependence of the absolute value of the Onsager phase factor $|\gamma|$ at $(U, V, T) = (0.5, 0.1, 1\times10^{-4})$.
(b) and (c) Illustration of the Berry curvature $B^z_{\nu,\sigma}({\bm k})$ and the energy band $E_{\nu,\sigma}({\bm k})$ near the Dirac point for two $\Delta$ values: $\Delta\ll|E_{\rm F}|$ and $\Delta>|E_{\rm F}|$.
}\label{Fig:PhaseFactor}
\end{centering}
\end{figure}
%
The $V'$-dependence of the absolute value of the Onsager phase factor $|\gamma|$ at $(U, V, T) = (0.5, 0.1, 1\times10^{-4})$ is plotted in Fig. \ref{Fig:PhaseFactor}.
The value of $|\gamma|$ becomes zero as $V'$ decreases, and increases to $0.5$ when $V'$ is sufficiently large.
This behavior is consistent with previously reported experimental results \cite{Kawasugi} in which the phase factor changes from $0.5$ to zero as the pressure $P$ decreases, where the $P$-dependence can be considered as the change in $V'$ associated with the change in $P$.
$|\gamma|$ does not become zero even when the system is in the interaction-induced QSH \tb{insulating phase} because of the following.
The Berry phase $\phi_B$ is calculated by the surface integral of the Berry curvature $B^z_{\nu,\sigma}({\bm k})$ on the Fermi surface $S_F$ as shown in Eq. (\ref{Eq:PhaseFactor}).
In a massive Dirac electron system such as an interaction-induced QSH \tb{insulating phase}, the peak of $B^z_{\nu,\sigma}({\bm k})$ decreases with the increase in the energy gap $\Delta$, and $B^z_{\nu,\sigma}({\bm k})$ widens and spreads in the Brillouin zone as shown in Fig. \ref{Fig:band-berrycurv} (b).
Therefore, when the spread of $B^z_{\nu,\sigma}({\bm k})$ becomes sufficiently larger than the integral range $S_F$ with the increase in $\Delta$ owing to $V'$ (see Figs. \ref{Fig:PhaseFactor} (b) and (c)), $\phi_B$ decreases and becomes zero ($|\gamma|$ becomes $0.5$).

%
\subsection{Temperature dependence of DC resistivity under ambient pressure}
Next, we calculated the $T$-dependence of the electronic state with the Hartree-Fock approximation and investigate temperature $T$ effects for the interaction-induced QSH insulating phase.
It has been reported that the DC resistivity of $\alpha$-(BETS)$_2$I$_3$ obtained experimentally \cite{Inokuchi, TajimaPriv, Kawasugi} was nearly constant at $T>50\times10^{-4}$ and sharply increases below $T=50\times10^{-4}$.
This experimental fact suggests that the electronic state changes around $T=50\times10^{-4}$ and a large band gap opens below this temperature.
To demonstrate the experimental results described above, we sweep $T$ as a parameter and investigate the $T$-dependence on DC conductivity using $T$-matrix approximation.
Throughout this subsection, $U=0.5$, $V=0.1$, and $V'=0.16$.

\begin{figure}
\begin{centering}
\includegraphics[width=87mm]{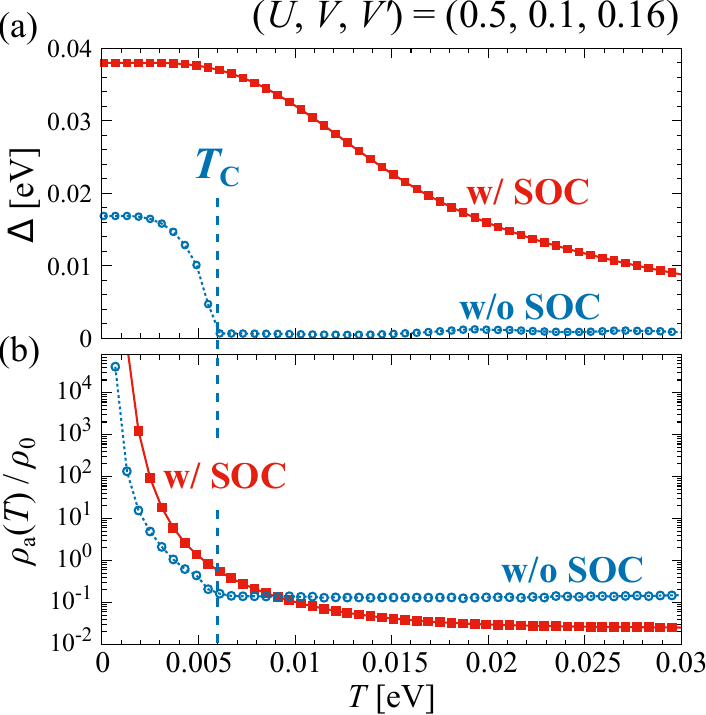}
\caption{(Color online)
$T$-dependence of (a) the energy gap $\Delta$ at $(U,V,V')=(0.5,0.1,0.16)$, and (b) the DC resistivity along the $a$-axis direction $\rho_a(T)/\rho_0$ in units of the universal conductivity reciprocal $\rho_0\equiv \sigma_0^{-1} = (4e^2/\pi h)^{-1}$ at $(U,V,V')=(0.5,0.1,0.16)$.
}\label{Fig:T-dependence}
\end{centering}
\end{figure}
Figure \ref{Fig:T-dependence} (a) shows the $T$-dependence of the energy gap $\Delta$ at $(U,V,V')=(0.5,0.1,0.16)$ with and without SOC.
When SOC is absent, the QAH \tb{insulating phase} in which \tb{the order parameter} $\chi^{\rm QAH}$ is finite and the Chern number $N^{\rm Ch}$ becomes $\pm1$ appears, and $\Delta$ without SOC increases sharply below the critical temperature $T_{\rm C}$ owing to the phase transition between the massless Dirac electron and QAH \tb{insulating phases}.
Conversely, when SOC is considered, $\Delta \sim 0.01$ at $T = 0.03$, which is approximately five times that of the case without the repulsive interaction.
On decreasing $T$, $\Delta$ with SOC gradually increases toward low-$T$ owing to the next-nearest-neighbor interaction-induced QSH \tb{insulating phase} and has a constant value $\Delta \sim 0.038$, which is twice that of the case without SOC.
This result indicates that the contribution of SOC and $V'$ renders the interaction-induced QSH \tb{insulating phase} more stable.
In Figs. \ref{Fig:T-dependence} (b), the $T$-dependence of the DC resistivity along the $a$-axis direction $\rho_a(T)/\rho_0\equiv\left(\sigma_a(T)/\sigma_0\right)^{-1}$ in units of the reciprocal of universal conductivity $\rho_0\equiv \sigma_0^{-1} = (4e^2/\pi h)^{-1}$ is plotted for two cases with and without SOC.
$\rho_a(T)/\rho_0$ without SOC is nearly constant at $T>T_{\rm C}$ and sharply increases at $T<T_{\rm C}$.
Alternatively, $\rho_a(T)/\rho_0$ with SOC increases continuously near $T=T_{\rm C}$ with the decrease in $T$ reflecting the gentle $T$-dependence on the energy gap $\Delta$.
Therefore, the sharp increase in DC resistivity below $T=50\times10^{-4}$ observed in experiments \cite{Inokuchi, TajimaPriv, Kawasugi} can be explained by considering the interaction-induced QSH insulating phase.
As mentioned in section III.B, the difference with and without SOC is analogous to a ferromagnet with and without an external magnetic field.

\begin{figure}
\begin{centering}
\includegraphics[width=85mm]{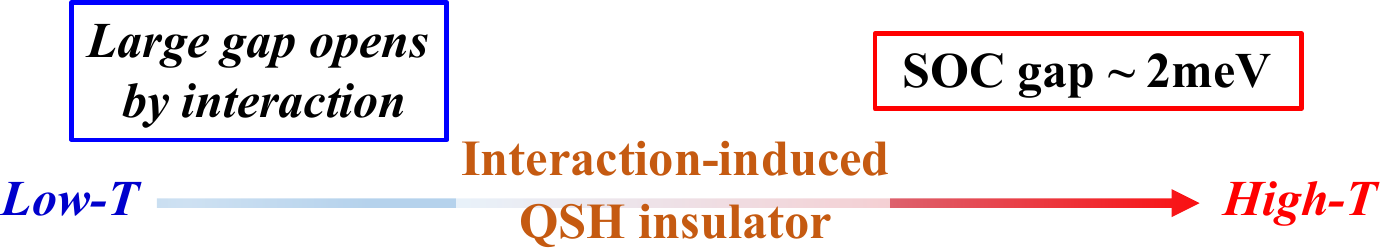}
\caption{(Color online)
Illustration of the $T$-dependence of the electronic state obtained by calculation with the Hartree-Fock approximation, with onsite $U$, nearest-neighbor sites $V$, and next-nearest-neighbor sites $V'$.
}\label{Fig:Schematic}
\end{centering}
\end{figure}
%
To summarize this subsection, we presented the $T$-dependence of the electronic state obtained with the Hartree-Fock approximation (Fig. \ref{Fig:Schematic}).
In this study, the interaction-induced QSH \tb{insulating phase} stably appears in the entire $T$-region.
As $T$ is decreased, the order parameter in the interaction-induced QSH \tb{insulating phase} $\chi^{\rm QSH}$ which is primarily caused by the nearest- and next-nearest-neighbor repulsions and SOC increases and the energy gap is gradually enhanced.
The interaction-induced QSH \tb{insulating phase} is consistent with experimental results such as X-ray diffraction, DC resistivity, and Shubnikov-de Haas oscillation \cite{KitouSawaTsumuraya, Inokuchi, TajimaPriv, Kawasugi} and is a strong candidate for the insulating state on $\alpha$-(BETS)$_2$I$_3$ at low-$T$.

\subsection{Spin fluctuations in the high-$T$ Dirac electron phase}
In this subsection, to investigate the effects of the repulsive interaction on spin fluctuations, we calculated the spin susceptibility using RPA and discuss the relationship with NMR experiments of $\alpha$-(BETS)$_2$I$_3$ \cite{Hiraki, Fujiyama, SekinePriv}.
Previous studies on $\alpha$-(ET)$_2$I$_3$ \cite{Kobayashi2013, HirataNat2016, Matsuno2017} have shown that the ferrimagnetic (FM) spin polarization observed in the site-resolved Knight shift \cite{HirataNat2016} is induced by $U$, where only the site-resolved Knight shift at B site $K_B$, defined in Eq. (\ref{eq:knight_shift}), becomes negative with the increase in $U$, and the other components always remain positive.
It has been reported that this behavior is caused by the inter-band electron-hole excitation enhanced by $U$ \cite{Kobayashi2013, HirataNat2016, Matsuno2017}.
In this subsection, we also investigated the possibility of FM spin polarization in $\alpha$-(BETS)$_2$I$_3$ \cite{SekinePriv}.

\begin{figure}
\begin{centering}
\includegraphics[width=85mm]{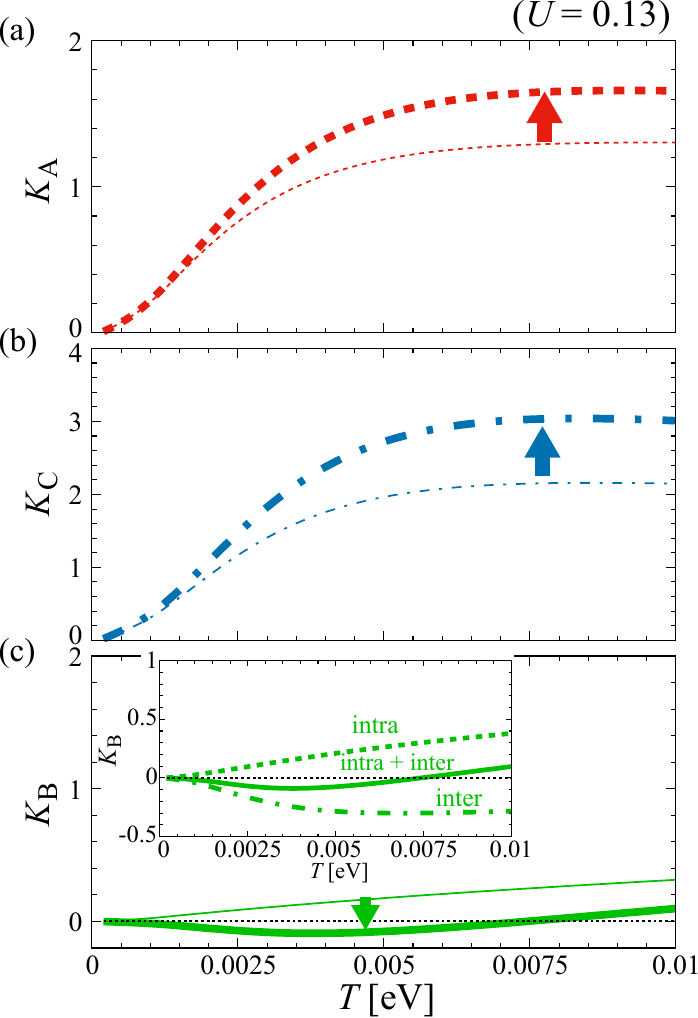}
\caption{(Color online)
$T$-dependence of the site-resolved Knight shift $K_\alpha \equiv \sum_{\beta} {\rm Re}\left[\chi^S_{\alpha,\beta}({\bm 0}, 0)\right]$ at $U = 0.13$ (thick line) and $U=0$ (thin line) for (a) $\alpha=$ A, (b) $\alpha=$ C, and (c) $\alpha=$ B.
The inset in (c) is the $T$-dependence of $K_{\rm B}$ divided into two components: intra-band (dotted line) and inter-band (one-dot chain line).
}\label{Fig:Knight_vs_T}
\end{centering}
\end{figure}
Figures \ref{Fig:Knight_vs_T}(a)-(c) show the $T$-dependence of the site-resolved Knight shift $K_\alpha \equiv \sum_{\beta} {\rm Re}\left[\chi^S_{\alpha,\beta}({\bm 0}, 0)\right]$, defined in Eq. (\ref{eq:knight_shift}) at the $\alpha = $ A, B, and C sites obtained using RPA at $U=0.13$ (thick line) and $U=0$ (thin line) as an example.
As RPA overestimates the magnitude of the repulsive interaction, we considered the $U$ value, which is smaller than those used in previous subsections ($U < 0.5$) for comparison with the experimental results.
As shown in Figs. \ref{Fig:Knight_vs_T} (a) and (b), $K_A$ and $K_C$ are enhanced when $U$ is considered, and become zero with the decrease in $T$ owing to the cancellation of each component of Re$\left[\chi^{S}_{{\rm A},\beta}\right]$ and Re$\left[\chi^{S}_{{\rm C},\beta}\right]$ ($\beta=$A, A$'$, B, and C).
In contrast, $K_B$ decreases and becomes negative below $T\sim 0.0075$ as shown in Fig.\ref{Fig:Knight_vs_T}(c).
The $T$-dependencies of Re$\left[\chi^{S,{\rm Intra}}_{{\rm B}}\right]$, Re$\left[\chi^{S,{\rm Inter}}_{{\rm B}}\right]$, and Re$\left[\chi^{S}_{{\rm B}}\right]$ are also plotted in the inset of Fig. \ref{Fig:Knight_vs_T} (c).
It is indicated that Re$\left[\chi^{S,{\rm Inter}}_{{\rm B}}\right]$ becomes negative for $0<T<0.01$, {and causes $Re\left[\chi^{S}_{{\rm B}}\right]$ to become negative}.
This behavior is qualitatively similar to that observed in $\alpha$-(ET)$_2$I$_3$\cite{HirataNat2016}.
This is because $\alpha$-(BETS)$_2$I$_3$ at high-$T$ has a characteristic wavenumber dependence on the square of the absolute value of the eigenvector such as a zero line for B and C sites, which is similar to that of $\alpha$-(ET)$_2$I$_3$ under high pressure.

\begin{figure}
\begin{centering}
\includegraphics[width=85mm]{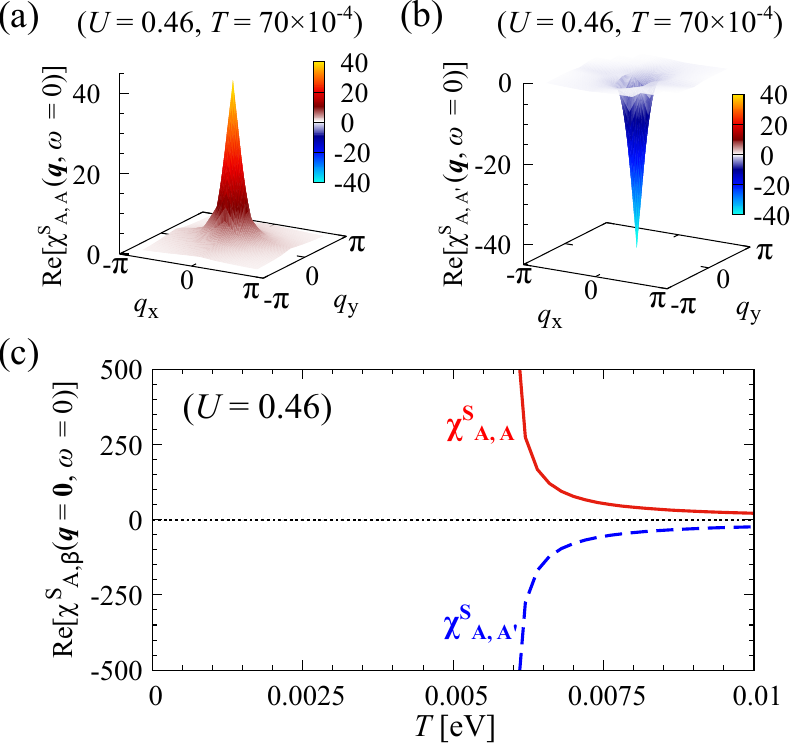}
\caption{(Color online)
Momentum ${\bm q}$-dependence of the real-part of the site-resolved spin susceptibility of the A site ${\rm Re}\left[\chi^S_{{\rm A},\beta}({\bm q}, 0)\right]$ at $(U,T)=(0.46, 0.007)$ for (a) $\beta = $ A and (b) $\beta = $ A$'$.
(c) $T$-dependence of ${\rm Re}\left[\chi^S_{{\rm A},{\rm A}}({\bm 0}, 0)\right]$ and ${\rm Re}\left[\chi^S_{{\rm A},{\rm A'}}({\bm 0}, 0)\right]$.
}\label{Fig:ReXSAbeta}
\end{centering}
\end{figure}
Next, we calculated spin susceptibility considering stronger interaction and investigated the type of spin susceptibility that was enhanced.
Figures \ref{Fig:ReXSAbeta}(a) and (b) show the momentum $\bm q$-dependence of the spin susceptibility Re$\left[\chi^S_{{\rm A},\beta}\right]$ at $\omega=0$ for $\beta={\rm A}, {\rm A}'$ with $(U, T)=(0.46, 0.007)$.
In the strong interaction case, Re$\left[\chi^S_{{\rm A},{\rm A}}\right]$ and Re$\left[\chi^S_{{\rm A},{\rm A}'}\right]$ exhibit a peak at ${\bm q}={\bm 0}$ reflecting the Fermi point in the Dirac electron system.
Re$\left[\chi^S_{{\rm A},{\rm A}}\right]$ and Re$\left[\chi^S_{{\rm A},{\rm A'}}\right]$ at ${\bm q}={\bm 0}$ have approximately the same absolute values with opposite signs.
Here, a negative peak in the site off-diagonal component Re$\left[\chi^S_{{\rm A},{\rm A}'}({\bm 0}, \omega=0)\right]$ indicates that there is spin fluctuation that aligns the spins between the A and A' sites in the unit cell to the opposite directions.
Furthermore, we plotted the $T$-dependence of Re$\left[\chi^S_{{\rm A},{\rm A}}({\bm q}={\bm 0},\omega=0)\right]$ and Re$\left[\chi^S_{{\rm A},{\rm A}'}({\bm q}={\bm 0},\omega=0)\right]$ (Fig. \ref{Fig:ReXSAbeta} (c)).
With the decrease in $T$, they are enhanced, and diverge positively and negatively toward $T\sim 0.006$.
This result indicates that the spin fluctuation inducing the antiferromagnetic (AF) spin order between A and A$'$ sites in the unit cell is enhanced as $T$ decreases, when a strong repulsive interaction is considered.
This AF spin ordered insulating phase corresponds to the spin ordered massive Dirac electron state mentioned in our previous study \cite{Ohki2020BETS}.

\begin{figure}
\begin{centering}
\includegraphics[width=85mm]{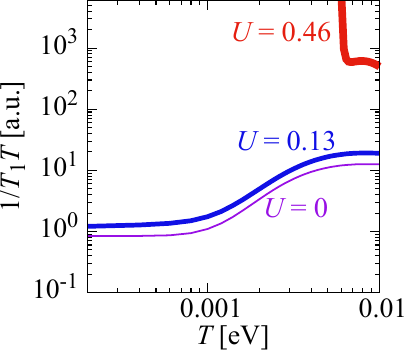}
\caption{(Color online)
$T$-dependence of $1/T_1T$ at $U=0$, $U=0.13$, and $U=0.46$.
}\label{Fig:T1T_vs_T}
\end{centering}
\end{figure}
Finally, the $T$-dependence of $1/T_1T=\sum_{\bm q}\sum_{\alpha}{\rm Im}\left[\chi^\pm_{\alpha,\alpha}({\bm q},\omega_0)\right]/\omega_0$ at $\omega_0=0.001$ is displayed in Fig. \ref{Fig:T1T_vs_T} for $U=0$, $U=0.13$ where FM spin fluctuation is dominant, and $U=0.46$ where AF spin fluctuation is dominant.
The FM spin polarization occurring at weak $U$ ($=0.13$) causes only a slight increase in the $1/T_1T$ value and has no significant contribution to $1/T_1T$.
However, the AF spin polarization causes a clear change in $1/T_1T$.
When a strong $U$ ($=0.46$) is considered and AF spin polarization occurs, $1/T_1T$ tends to increase and diverges as $T$ decreases.

\section{Summary and Discussion}

In conclusion, we have employed the candidate topological insulator $\alpha$-(BETS)$_2$I$_3$ as a model system and have proposed the realization of a QSH insulator that is induced by spin-orbit and repulsive interactions.
In the absence of SOC the repulsive interactions lead to two types of insulating ground states that are degenerate at mean-field level, namely, a QSH state and a QAH state without a magnetic field.
Mathematically, both of them have chiral edge states with a nonzero Chern number.
Switching on the spin-orbit interaction, a QSH insulator with helical edge states is favored over the QAH state, which is a cousin of the QSH state originally proposed for graphene under non-local repulsive interactions \cite{Raghu, Rachel}.
By constructing and carefully analyzing mean-field theory, we demonstrate that the insulating gap due to the spin-orbit interaction is drastically enhanced by the repulsions, leading to a QSH \tb{insulating phase} that is characterized by a nonzero spin Chern number and a finite order parameter.
Upon cooling or increasing the repulsive interactions, such insulating gap continuously grows and helps stabilizing the \tb{state} for realistic sizes of parameters.

In the mean-field studies for the honeycomb lattice QSH and QAH \tb{insulating phases} were proposed to appear when the next-nearest-neighbor repulsions ($V'$) were larger than the nearest-neighbor ones ($V$) \cite{Raghu}.
For a realistic condition ($V' < V$), however, subsequent studies using density matrix renormalization group technique \cite{Rachel,Wen2010,Weeks,Dauphin,Noel2013,Grushin2013,Daghofer2014,Duric,Capponi,Motruk,Tianhan,Venderbos} have failed to find such topological states. By contrast, as we discussed in this paper, long-range repulsive interactions in $\alpha$-(BETS)$_2$I$_3$ remain sizeable and can eventually generate an interaction-induced QSH \tb{insulating phase} even for such normal condition.
Of course, we cannot exclude the possibility that such QSH state may be wiped out if more sophisticated calculations above mean-field level is conducted.
However, the QSH state discussed here can explain a range of experimental findings in this material, such as the absence of inversion symmetry breaking, a metal-to-insulator crossover at low-$T$, and a topological Berry phase change under pressure \cite{Inokuchi, Kawasugi, TajimaPriv}.
Interestingly, a recent numerical study using a continuum Weyl model and incorporating long-range Coulomb interaction \cite{Hirata2021} have suggested several possible ordered phases in $\alpha$-(ET)$_2$I$_3$ -- a weak spin-orbit counterpart of $\alpha$-(ET)$_2$I$_3$ -- which vary from charge, spin, to bond ordered phases.
Because the theoretical frameworks for two systems would be almost identical (except for the size of spin-orbit interaction) and none of the above phases have been experimentally observed in $\alpha$-(BETS)$_2$I$_3$, the emergence of a QSH \tb{insulating phase} is at least consistent with \cite{Hirata2021} and would be rather conceivable.

Furthermore, we investigated the effect of spin fluctuations on the electronic state of $\alpha$-(BETS)$_2$I$_3$ using RPA \cite{Kobayashi2004, Kobayashi2013, HirataNat2016, Matsuno2017, Matsuno2018}.
For a weak interaction case, when $T$ decreased, only ferrimagnetic spin polarization appears owing to the characteristic wave function of $\alpha$-(BETS)$_2$I$_3$ \cite{SekinePriv}.
This behavior is similar to the ferrimagnetic spin polarization observed in $\alpha$-(ET)$_2$I$_3$ \cite{HirataNat2016}.

A recent NMR experiment reported that time reversal symmetry breaking was not observed, and $1/T_1T$ was proportional to the power of $T$ and varied continuously near $T=50\times10^{-4}$ \cite{Fujiyama}.
In our calculation using RPA, ferrimagnetic spin polarization has no significant effect on the $T$-dependence of $1/T_1T$ and this is consistent with the experimental results.
The ferrimagnetic spin polarization has been observed by a site-resolved NMR experiment in $\alpha$-(ET)$_2$I$_3$ \cite{HirataNat2016} and $\alpha$-(BETS)$_2$I$_3$ \cite{SekinePriv}.

It will be an exciting future study to examine the possibility of such QSH state within a continuum Dirac-Weyl model and consider other effects omitted in this study, such as orbital current and edge excitations.
In fact, calculations based on the Dirac Hamiltonian combined with $\beta$-detected NMR experiments in the topological insulator Bi$_{0.9}$Sb$_{0.1}$ have shown that the nuclear spin-lattice relaxation rate $1/T_1$ can be strongly affected by orbital currents and leads to a substantial change in the power of its temperature dependence \cite{Hirosawa, MacFarlane}.
Whether that kind of orbital effects show up in the QSH state in $\alpha$-(BETS)$_2$I$_3$ would be an interesting open question.
Possibilities of exotic higher-order topological states localizing at the intersection of edges of QSH insulators \cite{Hatsugai, Hashimoto, Benalcazar, Schindler2018Science, Kudo, Araki, Schindler2018Nature, Hohenadler2013, Rachel} should be also investigated by employing, for instance, a cylindrical boundary condition that also considers real-space textures.
Lastly, to address possible electron correlation effects above a mean-field level, just like in honeycomb lattice it would be informative to perform calculations using vertex corrections \cite{YoshimiMaebashi}, variational Monte Carlo method \cite{mVMC}, or functional renormalization group theory \cite{Tazai, Kontani}.
Such extended study will help us to gain deeper insight into the stability of the QSH state proposed here and its relationship to relevant models in the honeycomb lattice \cite{Rachel,Wen2010,Weeks,Dauphin,Noel2013,Grushin2013,Daghofer2014,Duric,Capponi,Motruk,Tianhan,Venderbos}.

\begin{acknowledgments}
The authors would like to thank S. Onari, Y. Yamakawa, and H. Kontani for the fruitful discussions.
We would also like to thank H. Sawa, T. Tsumuraya, and S. Kitou for their valuable comments.
We would like to express our gratitude to N. Tajima and Y. Kawasugi for informative discussions from the experimental aspects.
We would also like to express our gratitude to the referees for correcting the manuscript as well as reading and comments.
The computation in this work was performed using the facilities of the Supercomputer Center, Institute for Solid State Physics, University of Tokyo.
This work was supported by MEXT/JSPJ KAKENHI under grant numbers 21H01041, 19J20677, 19H01846, and 15K05166.
\end{acknowledgments}

\appendix
\renewcommand{\thefigure}{\Alph{section}.\arabic{figure}}
\setcounter{figure}{0}
\section{Values of transfer integrals and repulsive interactions}
In this appendix, we show the values of transfer integrals and repulsive interactions considering the screening effect using the cRPA method in the RESPACK code \cite{Nakamura}.
Throughout the interaction calculation, we set the energy cutoff of the dielectric function as 5.0 Ry.

\begin{table}
\centering
\caption{{List of transfer integrals without (with) SOC $t_{\alpha,\beta,\sigma}^{({\bm \delta})}$ (${t_{\alpha,\beta,\sigma}^{({\bm \delta})}}^{\rm SOC}$), and the effective repulsive interactions $W_{\alpha,\beta}^{({\bm \delta})}$ (units of meV).} {Here, sgn($\sigma$) is $+1 (\sigma=\uparrow)$ or $-1 (\sigma=\downarrow)$.}}\label{tab:TransWeff}
\begin{ruledtabular}
\begin{tabular}{llllll}
   \rule[-3.5mm]{0mm}{8mm}
    \hspace{0.2cm} & Re [$t_{\alpha,\beta,\sigma}^{({\bm \delta})}$] & Re [$t_{\alpha,\sigma;\beta,\sigma}^{({\bm \delta})}$]$^{\rm SOC}$ & Im [$t_{\alpha,\sigma;\beta,\bar{\sigma}}^{({\bm \delta})}$]$^{\rm SOC}$ & \hspace{0.2cm} &\\
    \colrule
   $t_{\rm a1}$ &\hspace{0.1cm} -10.12 &\hspace{0.1cm} -9.345 &\hspace{0.1cm} {\rm sgn}($\sigma$)$\times$1.365\\
   $t_{\rm a2}$ &\hspace{0.1cm} -16.31 &\hspace{0.1cm} -16.80 &\hspace{0.1cm} {\rm sgn}($\sigma$)$\times$0.206\\
   $t_{\rm a3}$ &\hspace{0.1cm} 51.08 &\hspace{0.1cm} 50.22 &\hspace{0.1cm} {\rm sgn}($\sigma$)$\times$0.614\\
   $t_{\rm b1}$ &\hspace{0.1cm} 138.1 &\hspace{0.1cm} 136.5 &\hspace{0.1cm} {\rm sgn}($\sigma$)$\times$12.06\\
   $t_{\rm b2}$ &\hspace{0.1cm} 158.7 &\hspace{0.1cm} 154.1 &\hspace{0.1cm} {\rm sgn}($\sigma$)$\times$19.46\\
   $t_{\rm b3}$ &\hspace{0.1cm} 65.84 &\hspace{0.1cm} 63.77 &\hspace{0.1cm} {\rm sgn}($\sigma$)$\times$8.866\\
   $t_{\rm b4}$ &\hspace{0.1cm} 18.65 &\hspace{0.1cm} 17.92 &\hspace{0.1cm} {\rm sgn}($\sigma$)$\times$4.205\\
   $t_{\rm a1}'$ &\hspace{0.1cm} 14.09 &\hspace{0.1cm} 13.88 &\hspace{0.1cm} {\rm sgn}($\sigma$)$\times$0.064 \\
   $t_{\rm a3}'$ &\hspace{0.1cm} 4.527 &\hspace{0.1cm} 4.425 &\hspace{0.1cm} 0.0 \\
   $t_{\rm a4}'$ &\hspace{0.1cm} 21.89 &\hspace{0.1cm} 21.68 &\hspace{0.1cm} 0.0 \\
   $t_{\rm b1}'$ &\hspace{0.1cm} -1.289 &\hspace{0.1cm} -1.276 &\hspace{0.1cm} {\rm sgn}($\sigma$)$\times$(-0.186) \\
   $t_{\rm b3}'$ &\hspace{0.1cm} -1.280 &\hspace{0.1cm} -1.161 &\hspace{0.1cm} 0.0 \\
   $t_{\rm b4}'$ &\hspace{0.1cm} 2.494 &\hspace{0.1cm} 2.636 &\hspace{0.1cm} 0.0 \\
   \hline  \hline
   \rule[-3.5mm]{0mm}{8mm}
   \hspace{0.15cm} &  Re [$W_{\alpha,\alpha}^{({\bm 0})}$] &\hspace{0.05cm} &\hspace{0.05cm} Re [$W_{\alpha,\beta}^{({\bm \delta})}$] &\hspace{0.1cm} &\hspace{0.1cm} \\
   \colrule
   $U_{\rm A}$ &\hspace{0.1cm} 1383 &\hspace{0.1cm} $V_{\rm a1}$ &\hspace{0.1cm} 580.5 &\hspace{0.1cm} \\
   $U_{\rm A'}$ &\hspace{0.1cm} 1383 &\hspace{0.1cm} $V_{\rm a2}$ &\hspace{0.1cm} 596.2 &\hspace{0.1cm} \\
   $U_{\rm B}$ &\hspace{0.1cm} 1396 &\hspace{0.1cm} $V_{\rm a3}$ &\hspace{0.1cm} 566.7 &\hspace{0.1cm} \\
   $U_{\rm C}$ &\hspace{0.1cm} 1359 &\hspace{0.1cm} $V_{\rm b1}$ &\hspace{0.1cm} 579.9 &\hspace{0.1cm} \\
   \hspace{0.1cm} &\hspace{0.1cm} &\hspace{0.1cm} $V_{\rm b2}$ &\hspace{0.1cm} 572.9 &\hspace{0.1cm} \\
   \hspace{0.1cm} &\hspace{0.1cm} &\hspace{0.1cm} $V_{\rm b3}$ &\hspace{0.1cm} 537.8 &\hspace{0.1cm} \\
   \hspace{0.1cm} &\hspace{0.1cm} &\hspace{0.1cm} $V_{\rm b4}$ &\hspace{0.1cm} 556.9 &\hspace{0.1cm} \\
   \hspace{0.1cm} &\hspace{0.1cm} &\hspace{0.1cm} $V_{\rm a1}'$ &\hspace{0.1cm} 329.8 &\hspace{0.1cm} \\
   \hspace{0.1cm} &\hspace{0.1cm} &\hspace{0.1cm} $V_{\rm a3}'$ &\hspace{0.1cm} 328.1 &\hspace{0.1cm} \\
   \hspace{0.1cm} &\hspace{0.1cm} &\hspace{0.1cm} $V_{\rm a4}'$ &\hspace{0.1cm} 329.1 &\hspace{0.1cm} \\
   \hspace{0.1cm} &\hspace{0.1cm} &\hspace{0.1cm} $V_{\rm b1}'$ &\hspace{0.1cm} 326.9 &\hspace{0.1cm} \\
   \hspace{0.1cm} &\hspace{0.1cm} &\hspace{0.1cm} $V_{\rm b3}'$ &\hspace{0.1cm} 323.9 &\hspace{0.1cm} \\
   \hspace{0.1cm} &\hspace{0.1cm} &\hspace{0.1cm} $V_{\rm b4}'$ &\hspace{0.1cm} 332.4 &\hspace{0.1cm} \\
\end{tabular}
\end{ruledtabular}
\end{table}
The values of the transfer integrals and repulsive interactions are listed in {Table \ref{tab:TransWeff}}.
The first column of the Table \ref{tab:TransWeff} presents the notations of the inter-molecular transfer integrals and interactions shown in Figs. \ref{Fig:2Dnetwork} (a) and (b).
The values of the real part of the transfer integrals without (with) SOC, Re $[t_{\alpha,\beta,\sigma}^{({\bm \delta})}]$ (Re $[t_{\alpha,\beta,\sigma}^{({\bm \delta})}]^ {\rm SOC}$), are listed in the second (third) column.
The fourth column shows the imaginary part of the transfer integrals with SOC Im $[t_{\alpha,\beta,\sigma}^{({\bm \delta})}]^ {\rm SOC}$.
The bottom of the second column lists the values of the real part of the effective direct integral Re $[W_{\alpha,\beta}^{({\bm \delta})}]$, calculated using RESPACK.
Here, ${\bm \delta}=(\delta_b,\delta_a)$ is the relative lattice vector in the $a$-$b$ plane, and $\alpha$ and $\beta$ are the molecule indices in the unit cell (A, A', B, C).
$W_{\alpha,\beta}^{({\bm \delta})}$ is an interaction taking into account the screening effect based on the first-principles calculations \cite{Nakamura}.
The average values of the nearest-neighbor and next-nearest-neighbor components are $V_{\rm a}=\frac{1}{3}\sum_{n=1}^3V_{{\rm a}n}=581.1$ meV, $V_{\rm b}=\frac{1}{4}\sum_{n=1}^4V_{{\rm b}n}=561.9$ meV, $V_{\rm a}'=\frac{1}{3}\sum_{n=1}^4V_{{\rm a}n}'=329.0$ meV, and $V_{\rm b}=\frac{1}{3}\sum_{n=1}^4V_{{\rm b}n}'=327.7$ meV.
As $V_b/V_a=0.967\simeq1$ and $V_b'/V_a'=0.996\simeq1$, there is charge geometrical frustration in $\alpha$-(BETS)$_2$I$_3$.
This charge geometrical frustration effect is one of characteristic features of organic conductors \cite{Yoshimi2020}.

\begin{figure}
\begin{centering}
\includegraphics[width=85mm]{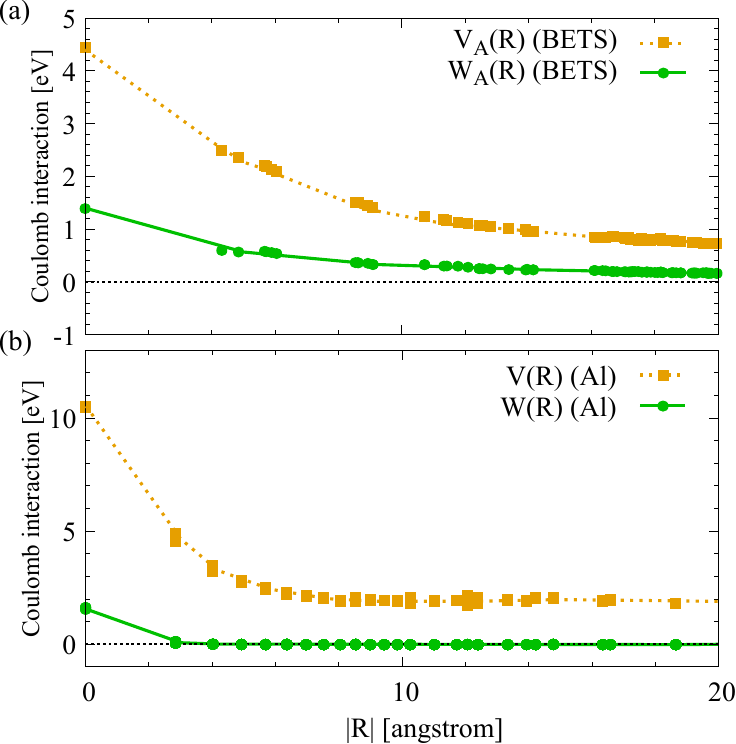}
\caption{(Color online) Range $|R|$-dependence of the bare direct integral $V_\alpha (R)$ and the static effective direct integral $W_\alpha (R)$ at the A site ($\alpha$ = A) for (a) $\alpha$-(BETS)$_2$I$_3$ at 30 K, and (b) result of aluminum (Al) as an example.
}\label{Fig:V-W_vs_R}
\end{centering}
\end{figure}
Figure \ref{Fig:V-W_vs_R} (a) shows the range($|R|$)-dependence of the bare direct integral $V_\alpha(R)$ estimated without the screening effect and the static effective direct integral $W_\alpha(R)$ for $\alpha={\rm A}$, evaluated using RESPACK.
It was confirmed that even when the screening effect is considered, the long-range components of $W_\alpha(R)$ have finite values.
Previous studies \cite{Nakamura2012, Misawa2020} have indicated that the values of the repulsive interaction obtained by the cRPA method decrease and the long-range repulsive interaction $W_\alpha(|R|\gg0)$ becomes zero when the dimensional down-folding method is used to reduce the three-dimensional Hamiltonian to two dimensions.
However, it is considered that an energy band of $\alpha$-(BETS)$_2$I$_3$ does not have a Fermi pocket because Dirac cones have an energy gap derived from the contribution of SOC and the Fermi energy is located to the center of this energy gap.
Therefore, the screening effect is expected to be weaker and it is more likely that the long-range components will survive.
Moreover, based on calculations using the Weyl model, it has been suggested that the long-range Coulomb interaction survives in Dirac electron systems, even when the screening effect is considered \cite{Khveshchenko, Kotov, Hirata2021}.
For comparison, we plotted $V_\alpha(R)$ and $W_\alpha(R)$ for aluminum in Fig. \ref{Fig:V-W_vs_R} (b).
In this case, the effective repulsive interaction is considerably decreased by the screening effect, and even the nearest-neighbor component becomes zero.

\setcounter{figure}{0}
\section{Relation of SOC and repulsive interactions}
In this appendix, we calculated the electronic state in the Hartree-Fock approximation and drew the phase diagram as a function of the strength of spin-orbit coupling and the onsite repulsion $U$ to compare with the results from preceding studies for other models such as the honeycomb lattice model \cite{Rachel2010, Rachel, RueggFiete2012}.
To change the strength of SOC as a parameter, the SOC values obtained by the first-principles calculation in this study is multiplied by $\lambda_{\rm SOC}$.
Furthermore, to investigate the contribution of $U$ to the electronic state in the presence of SOC, calculation was performed considering the spin order as a stable solution, which was prohibited in the main text.
It was also confirmed that the relation between parameters $(U, \lambda_{\rm SOC})$ and spin ordered massive Dirac electron (SMD) phase proposed in our preceding study \cite{Ohki2020BETS} which is caused by $U$ and associated with time-reversal symmetry breaking.

\begin{figure}
\begin{centering}
\includegraphics[width=75mm]{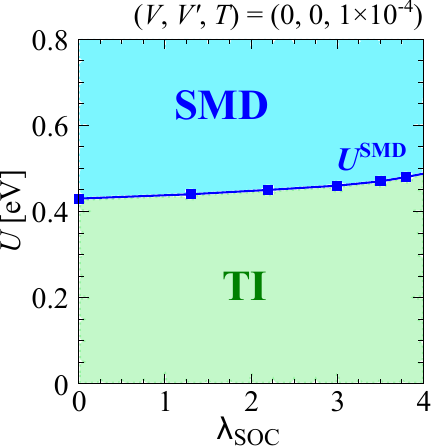}
\caption{(Color online) 
$U$-$\lambda_{\rm SOC}$ phase diagram for $(V, V', T) = (0, 0, 1\times10^{-4})$.
In $U < U^{\rm SMD}$, the topological insulator (TI) phase appears owing to the SOC contribution.
In $U > U^{\rm SMD}$, the spin ordered massive Dirac electron phase suggested in our previous study appears \cite{Ohki2020BETS}.
In the spin ordered massive Dirac electron phase, the time-reversal symmetry is broken owing to antiferromagnetism between the A and A$'$ sites in the unit cell.
}\label{Fig:U-lambda_SOC_PDG}
\end{centering}
\end{figure}
The calculation result is shown in Fig. \ref{Fig:U-lambda_SOC_PDG}.
As $U$ is increased and $U>U^{\rm SMD}$, spin ordered massive Dirac electron phase associated with antiferromagnetism between the A and A$'$ sites in the unit cell occurs \cite{Ohki2020BETS}.
In the spin ordered massive Dirac electron phase, the time-reversal symmetry is broken owing to antiferromagnetism between the A and A$'$ sites in the unit cell.
It is considered that this spin ordered phase corresponds to the AF insulator phase (AFI) in preceding studies for the honeycomb lattice model \cite{Rachel, Rachel2010, RueggFiete2012}.
In $U <U^{\rm SMD}$, the topological insulator (TI) phase appears owing to the contribution of SOC alone, as in the previous study for the honeycomb lattice model \cite{Rachel, Rachel2010, RueggFiete2012}.

\setcounter{figure}{0}
\section{Relation of spin ordered massive Dirac electron phase and the interaction-induced QSH \tb{insulating phase}}
Next, to investigate the relationship between the spin ordered massive Dirac electron phase in a previous study \cite{Ohki2020BETS} and the interaction-induced QSH \tb{insulating phase} and charge ordered insulating phases in $\alpha$-(BETS)$_2$I$_3$, calculation using the Hartree-Fock approximation with $U$, $V$, and $V'$ was performed by considering the spin order as a stable solution as in the previous appendix.
In previous studies for $\alpha$-(ET)$_2$I$_3$, a $U$-$V$ phase diagram is drawn and it has been shown that the spin order is stabilized when $U\gg0$ and $V$ is small, and the horizontal stripe charge order is stabilized when $V\gg0$ and $U$ is small \cite{Kobayashi2004, Kobayashi2004no2}.
In the following, we fixed $\lambda_{\rm SOC}$, $V$, and $T$ at $(\lambda_{\rm SOC}, V, T) = (1,0.1,1\times10^{-4})$ and draw a $U$-$V'$ phase diagram.

\begin{figure}
\begin{centering}
\includegraphics[width=80mm]{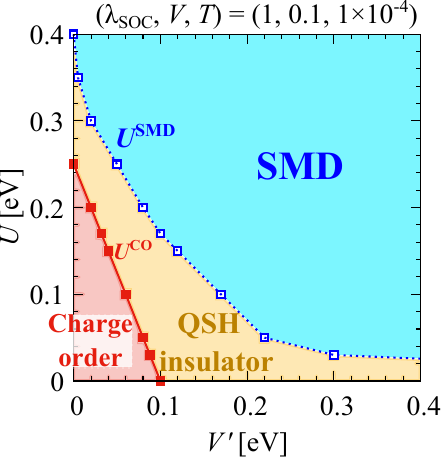}
\caption{(Color online)
$U$-$V'$ phase diagram for $(\lambda_{\rm SOC}, V, T) = (1, 0.1, 1\times10^{-4})$.
In $U<U^{\rm CO}$, the horizontal strip charge ordered insulating phase phase appears because $V$ is fixed at $V=0.1$.
In $U>U^{\rm SMD}$, the spin ordered massive Dirac electron phase appears owing to the contribution of $U$ \cite{Ohki2020BETS}.
The interaction-induced QSH \tb{insulating phase} appears during the transition between the spin ordered massive Dirac electron and horizontal strip charge ordered insulating phases, $U^{\rm CO}<U<U^{\rm SMD}$.
}\label{Fig:U-Vp_PDG}
\end{centering}
\end{figure}
Figure \ref{Fig:U-Vp_PDG} represents the $U$-$V'$ phase diagram calculation result.
When both $U$ and $V'$ are small such as $U<U^{\rm CO}$, the horizontal strip charge ordered insulating phase owing to the contribution of $V$ alone appears.
The spin ordered massive Dirac electron phase is stabilized in $U>U^{\rm SMD}$, and the interaction-induced QSH \tb{insulating phase} appears during the transition between the spin ordered massive Dirac electron and horizontal strip charge ordered insulating phases, $U^{\rm CO}<U<U^{\rm SMD}$.
This indicates that the interaction-induced QSH \tb{insulating phase} is not energetically stable under physically unrealistic values of $V$ and $V'$ such as in the case of $V'\gg V$ when the spin order is allowed as a stable solution in $\alpha$-(BETS)$_2$I$_3$, and not only $V'$ but also $V$ significantly contributes to the emergence of the interaction-induced QSH \tb{insulating phase}.

\nocite{*}

\bibliography{BETSQSH}

\end{document}